\documentclass[11pt,a4paper]{article}

\usepackage{amssymb}
\usepackage{amsmath}
\usepackage{mathrsfs}
\usepackage{graphicx}

\allowdisplaybreaks[1]

\DeclareMathOperator*{\tr}{tr}

\begin{document}

\thispagestyle{empty}

\begin{flushright}
 \large{KEK-TH-1344}
\end{flushright}

\vspace{10mm}
\begin{center}
{\LARGE \bf Quiver Chern-Simons Theories, D3-branes, \\
and Lorentzian Lie 3-algebras}

\vspace{10mm}
{\Large
Yoshinori Honma\footnote{E-mail address: 
yhonma@post.kek.jp}
and Sen Zhang\footnote{E-mail address: zhangsen@post.kek.jp}
}

\vspace{10mm}
 \large{ {\it Institute of Particle and Nuclear Studies, \\ High Energy Accelerator Research Organization (KEK) \\
 and \\ 
 Department of Particles and Nuclear Physics, \\
 The Graduate University for Advanced Studies (SOKENDAI), 
 \vspace{3mm} \\
Oho 1-1, Tsukuba, Ibaraki 305-0801, Japan}}

\end{center}

\vspace{20mm}
\ \

\begin{abstract}
\large{We show that the Bagger-Lambert-Gustavsson (BLG) theory with two pairs of negative norm generators is derived from the scaling limit of an orbifolded Aharony-Bergman-Jafferis-Maldacena (ABJM) theory. The BLG theory with many Lorentzian pairs is known to be reduced to the Dp-brane theory via the Higgs mechanism, so our scaling procedure can be used to derive Dp-branes directly from M2-branes in the field theory language. In this paper, we focus on the D3-brane case  and investigate the scaling limits of various quiver Chern-Simons theories obtained from different orbifolding actions. Remarkably, in the case of ${\cal{N}}=2$ quiver CS theories, the resulting D3-brane action covers a larger region  in the parameter space of the complex structure moduli than the ${\cal{N}}=4$ quiver CS theories. We also investigate how the $SL(2,Z)$ duality transformation is realized in the resultant D3-brane theory.}
\end{abstract}

\newpage
\setcounter{page}{1}
\setcounter{footnote}{0}

\baselineskip 6mm
\section{Introduction}
\setcounter{equation}{0}
Recently, there has been a lot of activities in superconformal Chern-Simons matter theories. They have arisen from searching the low energy
 effective action of multiple M2-branes. In \cite{Aharony:2008ug}, the action of an arbitrary number of multiple M2-branes was proposed by
 Aharony, Bergman, Jafferis, and Maldacena. It is an ${\cal{N}}=6$ superconformal $U(N) \times U(N)$ Chern-Simons matter theory, and the level of
 Chern-Simons term is $(k,-k)$. This ABJM theory has moduli space $Sym^N({\mathbb{C}}^4/{\mathbb{Z}}_k)$ and, therefore, is considered to
 describe $N$ M2-branes on an orbifold ${\mathbb{C}}^4/{\mathbb{Z}}_k$. On the other hand, triggered by the works of Bagger and Lambert \cite
{Bagger:2006sk} and Gustavsson \cite{Gustavsson:2007vu}, remarkable progress has also been achieved. The novelty is the appearance of new
 gauge structure, Lie 3-algebra. The BLG theory based on the Lie 3-algebra also has appropriate symmetries as the effective theory of multiple M2-branes, and under a particular realization of 3-algebra, the BLG theory actually coincides with the ABJM theory \cite{Bagger:2008se}. Furthermore, in \cite{Honma:2008jd} (see also \cite{Honma:2008ef,Antonyan:2008jf,Kluson:2009tz}), it was shown that the Lorentzian BLG (L-BLG)
 theory \cite{Gomis:2008uv,Benvenuti:2008bt,Ho:2008ei} based on the 3-algebra
\begin{align}
&[u,T^i,T^j]=f^{ij}_{ \ \ k}T^k , \ \ \ \ \ [T^i,T^j,T^k]=f^{ijk}v , \nonumber \\
&\textrm{tr} (u,v) =-1 , \ \ \ \ \ \textrm{tr} (T^i , T^j) = \delta^{ij}, \ \ \ \ ( u,v: \textrm{Lorentzian pair})
\label{LLie3}
\end{align}
 can be derived by taking a scaling limit of the ABJM theory. Because the L-BLG theory is reduced to the ordinary (2+1)d SYM via the Higgs
 mechanism, we can use this scaling procedure as a tool to obtain D2-branes directly from the ABJM theory in the field theory language.
The L-BLG theory was later generalized in \cite{Kobo:2009gz,Ho:2009nk,deMedeiros:2008bf} by involving additional pairs of negative norm
 generators.
 In \cite{Kobo:2009gz}, it was shown that this Extended L-BLG theory gives Dp-brane action whose worldvolume is compactified on torus $T^d \
 (d=p-2)$.
 Noting the fact that the Extended Lorentzian Lie 3-algebra can be regarded as the original 3-algebra (\ref{LLie3}) where the Lie algebra is replaced by
 the loop algebra, it is quite natural to expect that even the Extended L-BLG theory may be obtained from ABJM-like theory. Then, what type of
 model should we start from? The hint is given in \cite{Hashimoto:2008ij}. They showed that the D3-branes action can be derived from a particular
 quiver Chern-Simons theory obtained by orbifolding the ABJM action. Because the Extended L-BLG theory with two Lorentzian pairs is also
 reduced to the action of D3-branes through the Higgs mechanism, it is strongly expected that a certain scaling limit connecting the orbifolded
 ABJM theory and the Extended L-BLG theory exists.

In this paper, we show that the Extended L-BLG theory with two pairs of Lorentzian generators can be derived by taking a scaling
 limit of a ${\cal{N}}=4$ quiver Chern-Simons theory. This quiver CS theory describes M2-branes on ${\mathbb{C}}^4 / ({\mathbb{Z}}_{kn} \times
 {\mathbb{Z}}_n)$, and in our procedure, M2-branes are located very far from the origin of the orbifold.
 Taking $n \rightarrow \infty$ limit simultaneously, we make circle identifications in two directions, which are determined from the
${\mathbb{Z}}_{kn} , {\mathbb{Z}}_n$ orbifold actions. Our procedure corresponds to the ordinary $T^2$ compactification and this is why the
 Extended L-BLG theory emerges. This emergence has a useful application for obtaining the effective action of Dp-branes $(2 \leq p \leq 9)$
 from the ABJM theory using the Extended L-BLG theory. In this paper, we focus on the D3-brane case. We also investigate the scaling
 limit of various quiver CS theories obtained from different orbifoldings of the ABJM action.
 Moreover, we examine the $SL(2,Z)$ transformations after the reduction to the D3-brane theory and revisit the consideration given in
 \cite{Hashimoto:2008ij}. Remarkably, starting from the ${\cal{N}}=2$ quiver CS theories, the result is slightly different from the ${\cal{N}}=4$ case.
 In the ${\cal{N}}=4$ case, as in \cite{Hashimoto:2008ij}, the complexified coupling constant $\tau$ of the resultant D3-brane action depends on
 only one real parameter. However, in the ${\cal{N}}=2$ case, an additional degree of freedom appears, and therefore, we can cover a larger space of the
 complex structure moduli.

This paper is organized as follows. In section 2, we briefly review the BLG theory and its generalization. Then, we take a quick look at the
 ABJM theory, its scaling limit, and a ${\cal{N}}=4$ quiver Chern-Simons theory obtained by using the ordinary orbifold projection to the ABJM
 theory. In section 3, we explicitly show how to derive the Extended Lorentzian BLG theory with two Lorentzian pairs from a scaling limit of a
 ${\cal{N}}=4$ quiver CS theory and investigate the constraint on the $T^2$ compactification. Furthermore, in section 4, we apply our scaling limit
 to several quiver CS theories obtained by different ${\mathbb{Z}}_n$ orbifoldings. In section 5, we investigate the realization of $SL(2,Z)$
 transformations of the resultant D3-brane theory. Finally, we conclude in section 6.

\section{Effective theories of M2-branes}
\setcounter{equation}{0}
\subsection{BLG theory and its generalization}
We first provide a brief review of the BLG theory and its generalization. The BLG theory is a three dimensional conformal field theory with
 ${\cal{N}}=8$ supersymmetry. It contains 8 real scalar fields $X^I=\sum_a X^I_a T^a \ (I=1,\cdots,8)$, 
gauge fields $A^{\mu}=\sum_{a,b} A^{\mu}_{ab}T^a \otimes T^b \ (\mu=0,1,2)$ with two gauge indices, and a 16-component Majorana spinor
 field $\psi=\sum_a \psi_a T^a$. 

The Lagrangian of the BLG theory is given by
\begin{align}
L=-\frac{1}{2}\textrm{tr}(D^{\mu}X^I , D_{\mu}X^I)+\frac{i}{2}\textrm{tr}(\bar{\psi},\Gamma^{\mu}D_{\mu}\psi )
+\frac{i}{4}\textrm{tr}(\bar{\psi},\Gamma_{IJ}[X^I,X^J,\psi])-V(X)+L_{CS},
\label{BLGaction}
\end{align}
where $[T^a,T^b,T^c]=f^{abc}_{ \ \ \ d}T^d$ and the covariant derivative is defined by
\begin{align}
(D_{\mu}X^I)_a=\partial_{\mu}X^I_a-f^{cdb}_{ \ \ \ a}A_{\mu c d}(x)X^I_b.
\end{align}
$V(X)$ is a sextic potential term
\begin{align}
V(X)=\frac{1}{12}\textrm{tr}([X^I,X^J,X^K],[X^I,X^J,X^K]),
\end{align}
and the Chern-Simons term is given by
\begin{align}
L_{CS}=\frac{1}{2}\epsilon^{\mu\nu\lambda} \textrm{tr} \left( f^{abcd}A_{\mu ab}\partial_{\nu}A_{\lambda cd}+\frac{2}{3}f^{cda}_{ \ \ \
 g}f^{efgb}A_{\mu ab}A_{\mu cd}A_{\lambda ef} \right) .
\end{align}
Note that the level of the Chern-Simons term is chosen to be $k=1$ for simplicity.

In \cite{Kobo:2009gz} (see also \cite{Ho:2009nk,deMedeiros:2008bf}), the Lorentzian BLG theory based on the 3-algebra (\ref{LLie3}) was generalized by adding $d$ pairs of negative norm generators.
 Then, they showed that the worldvolume theory of Dp-branes ($p=d+2$) is produced. The proposed 3-algebra is
\begin{align}
&[u_0,u_a,u_b]=0, \nonumber \\
&[u_0,u_a,T^i_{\vec{m}}]=-im_a T^i_{\vec{m}}, \nonumber \\
&[u_0,T^i_{\vec{m}},T^j_{\vec{n}}]=im_a v^a \delta_{\vec{m}+\vec{n}}\delta^{ij}+f^{ij}_{ \ \ k}T^k_{\vec{m}+\vec{n}}, \nonumber \\
&[T^i_{\vec{l}},T^j_{\vec{m}},T^k_{\vec{n}}]=f^{ijk}\delta_{\vec{l}+\vec{m}+\vec{n}}v^0,
\label{ELLie3}
\end{align}
where $a,b=1,\cdots,d$ and $\vec{l},\vec{m},\vec{n} \in {\mathbb{Z}}^d$. $a$ and $b$ correspond to the label of the compactified
 direction and $\vec{m}$ to the Kaluza-Klein momentum\footnote{Instead, we can consider $\vec{m}$ as the index describing open string modes
 that interpolate the mirror images of a point in $S^1={\mathbb{R}}/{\mathbb{Z}}$ in the spirit of Taylor's T-duality \cite{Taylor:1996ik}.} along the $T^d$.
 $f^{ijk} \ (i,j,k=1, \cdots,{\textrm{dim}} \ {\mathbf{g}})$ is a structure constant of an arbitrary Lie algebra $\mathbf{g}$.
This 3-algebra actually satisfies the fundamental identity.
The nonvanishing part of the metric is
\begin{align}
\textrm{tr}( u_A,v^B ) =-\delta^B_A, \ \ \ \textrm{tr}( T^i_{\vec{m}},T^j_{\vec{n}} ) =\delta^{ij}\delta_{\vec{m}+\vec{n}}. \ \ \ (A=0,1,\cdots ,d)
\end{align} 

Following \cite{Kobo:2009gz}, we will rewrite the BLG action (\ref{BLGaction}) and derive the action of Dp-branes ($p=d+2$).
 The steps are summarized as follows. 
First, we derive 3d ${\cal{N}}=8$ SYM through the Higgs mechanism \cite{Mukhi:2008ux}. The difference from the original L-BLG theory is
 that the resulting D2-brane action has a Kaluza-Klein tower. Then, we obtain the Dp-brane action with a rearrangement of fields
 corresponding to T-duality. The worldvolume of Dp-brane is given as a flat $T^d$ bundle over the membrane worldvolume $\cal{M}$.

In the remainder of this subsection, we look at the above procedure more explicitly.
 For the 3-algebra (\ref{ELLie3}), we expand the fields as
\begin{align}
X^I&=X^I_{(i\vec{m})}T^i_{\vec{m}}+X^{IA}u_A+\underline{X}^I_Av^A, \nonumber \\
\psi &= \psi_{(i\vec{m})}T^i_{\vec{m}}+\psi^A u_A+\underline{\psi}_A v^A, \nonumber \\
A_{\mu}&=A_{\mu (i\vec{m})(j\vec{n})}T^i_{\vec{m}} \wedge T^j_{\vec{n}}+\frac{1}{2}A_{\mu(i\vec{m})}u_0 \wedge T^i_{\vec{m}}+\frac{1}{2}
A^a_{\mu(i\vec{m})}u_a \wedge T^i_{\vec{m}} \nonumber \\
& \ \ \ \ \ +\frac{1}{2}A^a_{\mu}u_0 \wedge u_a +A^{ab}_{\mu}u_a \wedge u_b + \textrm{(terms including $v^A$)}.
\end{align}

Each bosonic component has the following role:
\begin{itemize}
\item $X^I_{(i \vec{m})}$ : These fields become scalar fields corresponding to the transverse coordinates of Dp-branes and gauge fields along
 the fiber direction.

\item $X^{IA}$ : Higgs fields whose VEVs determine the moduli of $T^d$ and the circle radius in the M-direction.

\item $\underline{X}^I_A$ : Ghost fields that can be removed by Higgs mechanism.

\item $A_{\mu (i \vec{m})}$ : Gauge fields along ${\cal{M}}$.
\end{itemize}
The other bosonic terms do not show up in the following discussion.

Because the ghost fields $\underline{X}$ and $\underline{\psi}$ appear linearly in the action, these fields become Lagrange multipliers 
and can be integrated out. This gives constraint equations for $X^{IA}$ and $\psi^A$:
\begin{align}
\partial^{\mu}\partial_{\mu}X^{IA}=0, \ \ \ \Gamma^{\mu}\partial_{\mu}\psi^A=0.
\end{align}
As a solution, we choose a constant vector $\vec{X}^A=\vec{\lambda}^A$ and it
 determines the (d+1)-dimensional subspace ${\mathbb{R}}^{d+1} \subset {\mathbb{R}}^8$.
 ${\mathbb{R}}^{d+1}$ is compactified on $T^{d+1}$ and VEVs $\vec{\lambda}^{IA}$ give the moduli of the $T^d$ compactification and the M-theory circle. We can represent the metric of torus $T^d$ as
\begin{align}
G^{AB}=\vec{\lambda}^A\cdot \vec{\lambda}^B.
\label{met}
\end{align}

The covariant derivative becomes
\begin{align}
(D_{\mu}X^I)_{(i\vec{m})}=(\hat{D}_{\mu}X^I)_{(i\vec{m})}-A'_{\mu(i\vec{m})}\lambda^{I0}-im_aA_{\mu(i\vec{m})}\lambda^{Ia},
\end{align}
where
\begin{align}
(\hat{D}_{\mu}X^I)_{(i\vec{m})}&=\partial_{\mu}X^I_{(i\vec{m})}-f^{jk}_{ \ \ i}A_{\mu(k\vec{n})}X^I_{(j,\vec{m}-\vec{n})}, \nonumber \\
A'_{\mu(i\vec{m})}&=-im_a A^a_{\mu(i\vec{m})}+f^{jk}_{ \ \ i}A_{\mu(j,\vec{m}-\vec{n})(k\vec{n})}.
\end{align}

The Chern-Simons term is written as
\begin{align}
L_{CS}&=\frac{1}{2} A'_{(i\vec{m})} \wedge F_{(i,-\vec{m})}+\textrm{(total derivative)},
\end{align}
where $F_{\mu\nu (i,\vec{m})}=\partial_{\mu}A_{\nu(i\vec{m})}-\partial_{\nu}A_{\mu(i\vec{m})}-f^{jk}_
{ \ \ i}A_{\mu(j\vec{n})}A_{\nu(k,\vec{m}-\vec{n})}.$
Integrating $A'_{(i\vec{m})}$, Chern-Simons gauge fields obtain a degree of freedom and the usual $F^2$ term emerges.

The bosonic potential term is given by the square of a triple product
\begin{align}
[X^I,X^J,X^K]_{(i\vec{m})}=-im_a \lambda^{[I0}\lambda^{Ja}X^{K]}_{(i\vec{m})}+f^{jk}_{ \ \ i}\lambda^{[I0}X^J_{(j\vec{n})}X^{K]}_{(k, \vec{m}-
\vec{n})}.
\end{align}
The square of this term gives
\begin{align}
&6g^{ab}m_a m_b X^I_{\vec{m}}P^{IJ}_{\vec{m}}X^J_{-\vec{m}}-i\lambda^{[I0}\lambda^J_{\vec{m}}X^{K]}_{(i\vec{m})} f^{jk}_{ \ \ i}\lambda^{[I0}X^J_{(j\vec{n})}X^{K]}_{(k,-\vec{m}-\vec{n})} \nonumber \\
& \ -3 \Big[ G^{00} \langle [X^J,X^K]^2 \rangle -2 \langle [(\vec{\lambda}^0 \cdot \vec{X}) , X^I]^2 \rangle \Big],
\label{ELbos}
\end{align}
where
\begin{align}
&P^{IJ}_{\vec{m}} \equiv \delta^{IJ}-\frac{|\vec{\lambda}^0|^2 \lambda^I_{\vec{m}}\lambda^J_{\vec{m}} +
|\lambda_{\vec{m}}|^2 \lambda^{I0}\lambda^{J0}
-(\vec{\lambda}^0 \cdot \vec{\lambda}_{\vec{m}})(\lambda^{I0}\lambda^J_{\vec{m}}+\lambda^{J0}\lambda^I_{\vec{m}})}
{|\vec{\lambda}^0|^2|\vec{\lambda}_{\vec{m}}|^2-(\vec{\lambda}^0 \cdot \vec{\lambda}_{\vec{m}})^2}, \nonumber \\
&\vec{\lambda}_{\vec{m}} \equiv m_a\vec{\lambda}^a.
\end{align}

By collecting all the results, we obtain the D2-brane action with Kaluza-Klein tower. Then, we decompose $X^I$ as
\begin{align}
X^I=P^{IJ}X^J+\frac{1}{G^{00}}\lambda^{I0}(\vec{\lambda}^0 \cdot \vec{X})+\left( -\frac{G^{0a}}{G^{00}}\lambda^{I0}+\lambda^{Ia} \right),
\label{Xdec}
\end{align}
and regard the Kaluza-Klein masses $m_a$ with the derivatives of fiber direction $-i\partial_a$, we obtain the kinetic term of the fiber direction and the interaction term in the language of the Dp-brane worldvolume.

As a result, we obtain the following standard Dp-brane action\footnote{The tilde indicates that the fields are (3+d)-dimensional:
 $\tilde{\Phi}(x,y)=\sum_{\vec{m}}\Phi_{\vec{m}}(x)e^{i\vec{m}\cdot \vec{y}}$. $P^{IJ} \equiv \delta^{IJ}-\lambda^{IA}\pi^J_A$ is a 
projector into the subspace orthogonal to all $\vec{\lambda}^A$, where $\vec{\pi}_A$ is a dual basis satisfying $\vec{\lambda}^A \cdot
 \vec{\pi}_B=\delta^A_B$.} 
\begin{align}
L_{Dp}&=L_A+L_{F\tilde{F}}+L_{X}+L_{pot}, \nonumber \\
L_A&=-\frac{1}{4G^{00}}\int \frac{d^d y}{(2\pi )^d}\sqrt{g} \ (\tilde{F}^2_{\mu\nu}+2g^{ab}\tilde{F}_{\mu a}\tilde{F}_{\mu b}
+g^{ac}g^{bd}\tilde{F}_{ab}\tilde{F}_{cd}), \nonumber \\
L_{F\tilde{F}}&=\frac{G^{0a}}{8G^{00}}\int \frac{d^dy}{(2\pi )^d}\sqrt{g} \ (4\epsilon^{\mu\nu\lambda} \tilde{F}_{\mu a}\tilde{F}_{\nu\lambda}),
\nonumber \\
L_X&=-\frac{1}{2}\int \frac{d^dy}{(2\pi )^d}\sqrt{g} \ (\hat{D}_{\mu} \tilde{X}^I P^{IJ} \hat{D}_{\mu} \tilde{X}^J+g^{ab}\hat{D}_a\tilde{X}^I
P^{IJ}\hat{D}_b\tilde{X}^J), \nonumber \\
L_{pot}&=\frac{G^{00}}{4} \int \frac{d^d y}{(2\pi )^d} \sqrt{g}[P^{IK}\tilde{X}^K,P^{JL}\tilde{X}^L]^2,
\end{align}
whose worldvolume is ${\cal{M}} \times T^d$ with the metric
\begin{align}
ds^2=\eta_{\mu\nu}dx^{\mu}dx^{\nu}+g_{ab}dy^a dy^b,
\end{align}
where $g_{ab}=(G^{00}G^{ab}-G^{a0}G^{b0})^{-1}$ is the metric of dual torus.
\subsection{Orbifolding the ABJM theory}
The ABJM theory is a 3d ${\cal{N}}=6$ $U(N) \times U(N)$ Chern-Simons matter theory. This theory is conjectured to describe the low energy
 physics of $N$ M2-branes probing ${\mathbb{C}}^4/{\mathbb{Z}}_k$. The bosonic action of the ABJM theory is given by
\begin{align}
S=\int d^3x \Big[ &-\textrm{tr} \{ (D_{\mu}Z^A)^{\dagger}D^{\mu}Z^A+(D_{\mu}W^A)^{\dagger}D^{\mu}W^A \} -V(Z,W) \nonumber \\
&+\frac{k}{4\pi}\epsilon^{\mu\nu\lambda} \textrm{tr} \Big( A^{(1)}_{\mu}\partial_{\nu}A^{(1)}_{\lambda}+\frac{2i}{3}A^{(1)}_{\mu}A^{(1)}_{\nu}A^{(1)}_{\lambda} \nonumber \\
& \ \ \ \ \ \ \ \ \ \ \ \ \ \ \ \ \ \ \ \ \ -A^{(2)}_{\mu}\partial_{\nu}A^{(2)}_{\lambda}-\frac{2i}{3}A^{(2)}_{\mu}A^{(2)}_{\nu}A^{(2)}_{\lambda} \Big) \Big],
\end{align}
where $A=1,2$. $Z^A$ and $W^{\dagger A}$ are bifundamental matter fields and their covariant derivatives are defined by
\begin{align}
D_{\mu}Z^A&=\partial_{\mu}Z^A+iA^{(1)}_{\mu}Z^A-iZ^A A^{(2)}_{\mu}, \nonumber\\
D_{\mu}W^A&=\partial_{\mu}W^A+iA^{(2)}_{\mu}W^A-iW^A A^{(1)}_{\mu}.
\end{align}

In \cite{Honma:2008jd}, we explicitly show that the original L-BLG theory based on (\ref{LLie3}) is derived from the ABJM theory. Motivated by
 the agreement of the gauge structure of these two theories through the In\"{o}n\"{u}-Wigner contraction, we performed the following rescaling:
\begin{align}
&Z^A_0 \rightarrow \lambda^{-1}Z^A_0, \nonumber \\
&W^A_0 \rightarrow \lambda^{-1} W^A, \nonumber \\
&B_{\mu} \equiv (A^{(1)}_{\mu}-A^{(2)}_{\mu})/2 \rightarrow \lambda B_{\mu}, \nonumber \\
&k \rightarrow \lambda^{-1}k,
\end{align}
to the ABJM theory and took the $\lambda \rightarrow 0$ limit, where $Z^A_0$ and $W^A_0$ are the VEV of $Z^A$ and $W^A$. Then, we
 obtained the action of the L-BLG theory. This scaling limit corresponds to locate the M2-branes very far from the origin of the ${\mathbb{Z}}_k$
 orbifold so as not to feel the singularity and simultaneously take $k \rightarrow \infty$. Thus, this procedure is effectively the same as the ordinary
 $S^1$ compactification and that is why we obtain the L-BLG theory, which is almost D2-branes theory.

As explained in \cite{Kobo:2009gz}, the Extended Lorentzian 3-algebra (\ref{ELLie3}) can be regarded as the original Lorentzian 3-algebra with
 a loop algebra. Thus, it is natural to presume that even the Extended L-BLG theory might be derived from an M2-brane theory in a certain scaling
 limit. So which M2-brane theory is appropriate?
 In \cite{Hashimoto:2008ij}, it was shown that the D3-brane action can be derived by orbifolding the ABJM theory and taking a limit.
 Because the Extended L-BLG theory with $d=1$ also reduces to the D3-brane theory via the Higgs mechanism, these two theories might be connected
 directly. The main purpose of this paper is to clarify the relationship between the orbifolded ABJM theory and the Extended L-BLG theory.

In the remainder of this section, we review the orbifolded ABJM action. By applying the standard orbifolding technique \cite{Douglas:1996sw} to the ABJM theory or 
alternatively using the brane construction, we can derive various quiver Chern-Simons matter theories\footnote{For M2-branes on more
 general backgrounds, see \cite{Martelli:2008si,Hanany:2008cd,Fuji:2008yj,Ueda:2008hx,Hanany:2008fj,Franco:2008um,Franco:2009sp,Davey:2009sr,Aganagic:2009zk,
Taki:2009wf,Benini:2009qs} for
 example.} \cite{Benna:2008zy,Imamura:2008nn,Imamura:2008dt}. Here, we see a particular 3d
 ${\cal{N}}=4$ theory whose bosonic action is\footnote{This is the ``non-chiral orbifold gauge theory" described in \cite{Benna:2008zy} and we
 use their notation. This theory can also be regarded as case II in \cite{Terashima:2008ba} and the $n_A=n_B$ case in \cite{Imamura:2008nn} with alternate NS5- and (k,1)5-branes.
 The ``generalized ABJM model" described in \cite{Hashimoto:2008ij} is obtained by interchanging our $Z^{(2l)}$ and $W^{(2l)}$ in (\ref{N4QCS}).}
\begin{align}
 S=\int d^3 x \Big[ &-\textrm{tr}\sum^{2n}_{s=1} \{ (D_{\mu}Z^{(s)})^{\dagger}D^{\mu}Z^{(s)}+(D_{\mu}W^{(s)})^{\dagger}D^{\mu}W^{(s)} \} -V_{bos} \nonumber \\
&+\frac{k}{4\pi}\epsilon^{\mu\nu\lambda} \sum^n_{l=1} \textrm{tr} \{A^{(2l-1)}_{\mu}\partial_{\nu}A^{(2l-1)}_{\lambda}+\frac{2i}{3}A^{(2l-1)}_{\mu}A^{(2l-1)}_{\nu}A^{(2l-1)}_{\lambda} \nonumber \\
& \ \ \ \ \ \ \ \ \ \ \ \ \ \ \ \ \ \ \ \ \ -A^{(2l)}_{\mu}\partial_{\nu}A^{(2l)}_{\lambda}-\frac{2i}{3}A^{(2l)}_{\mu}A^{(2l)}_{\nu}A^{(2l)}_{\lambda} \} \Big].
\label{N4QCS}
\end{align}

\begin{figure}[h]
\centering
\includegraphics[width=12cm]{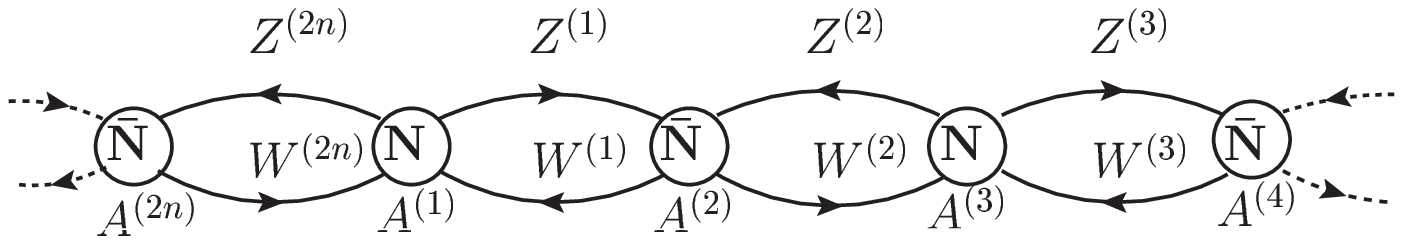}
\caption{Quiver diagram for ${\cal{N}}=4$ quiver CS theory (\ref{N4QCS}). This theory has global $SU(2)_o \times SU(2)_e$ symmetry and the $SU(2)_o$
 part rotates the fields on the odd links and the $SU(2)_e$ part corresponds to the even links.}
\label{N=4}
\end{figure}

The explicit forms of the covariant derivatives and bosonic potential are given by
\begin{align}
D_{\mu} Z^{(2l-1)} &=\partial_{\mu}Z^{(2l-1)}+iA^{(2l-1)}_{\mu}Z^{(2l-1)}-iZ^{(2l-1)}A^{(2l)}_{\mu}, \nonumber \\
D_{\mu} Z^{(2l)} &=\partial_{\mu}Z^{(2l)}+iA^{(2l+1)}_{\mu}Z^{(2l)}-iZ^{(2l)}A^{(2l)}_{\mu},\nonumber \\
D_{\mu} W^{(2l-1)} &=\partial_{\mu}W^{(2l-1)}+iA^{(2l)}_{\mu}W^{(2l-1)}-iW^{(2l-1)}A^{(2l-1)}_{\mu}, \nonumber \\
D_{\mu} W^{(2l)} &=\partial_{\mu}W^{(2l)}+iA^{(2l)}_{\mu}W^{(2l)}-iW^{(2l)}A^{(2l+1)}_{\mu},
\label{coder1}
\end{align}
\begin{align}
V_{bos}=&-\frac{4 {\pi}^2}{3 k^2}\sum^n_{l=1} \Big[ \ 
\textrm{tr} Y^A_{2l}Y^{\dagger}_{A,2l}Y^B_{2l}Y^{\dagger}_{B,2l}Y^C_{2l}Y^{\dagger}_{C,2l}+3\textrm{tr} Y^A_{2l}Y^{\dagger}_{A,2l}Y^B_{2l}Y^{\dagger}_{B,2l}Y^C_{2l+1}Y^{\dagger}_{C,2l+1} \nonumber \\
&+3\textrm{tr} Y^A_{2l}Y^{\dagger}_{A,2l}Y^B_{2l+1}Y^{\dagger}_{B,2l+1}Y^C_{2l+1}Y^{\dagger}_{C,2l+1}+\textrm{tr} Y^A_{2l+1}Y^{\dagger}_{A,2l+1}Y^B_{2l+1}Y^{\dagger}_{B,2l+1}Y^C_{2l+1}Y^{\dagger}_{C,2l+1} \nonumber \\
\nonumber \\
&+\textrm{tr} Y^{\dagger}_{A,2l-1}Y^A_{2l-1}Y^{\dagger}_{B,2l-1}Y^B_{2l-1}Y^{\dagger}_{C,2l-1}Y^C_{2l-1}+3\textrm{tr} Y^{\dagger}_{A,2l-1}Y^A_{2l-1}Y^{\dagger}_{B,2l-1}Y^B_{2l-1}Y^{\dagger}_{C,2l}Y^C_{2l} \nonumber \\
&+3\textrm{tr} Y^{\dagger}_{A,2l-1}Y^A_{2l-1}Y^{\dagger}_{B,2l}Y^B_{2l}Y^{\dagger}_{C,2l}Y^C_{2l}+\textrm{tr} Y^{\dagger}_{A,2l}Y^A_{2l}Y^{\dagger}_{B,2l}Y^B_{2l}Y^{\dagger}_{C,2l}Y^C_{2l} \nonumber \\
\nonumber \\
&+4\textrm{tr} Y^A_{2l-1}Y^{\dagger}_{B,2l-1}Y^C_{2l-1}Y^{\dagger}_{A,2l-1}Y^B_{2l-1}Y^{\dagger}_{C,2l-1}+12\textrm{tr} Y^A_{2l}Y^{\dagger}_{B,2l}Y^C_{2l+1}Y^{\dagger}_{A,2l+2}Y^B_{2l+2}Y^{\dagger}_{C,2l+1} \nonumber \\
&+12\textrm{tr} Y^A_{2l+1}Y^{\dagger}_{B,2l+1}Y^C_{2l}Y^{\dagger}_{A,2l-1}Y^B_{2l-1}Y^{\dagger}_{C,2l}+4\textrm{tr} Y^A_{2l}Y^{\dagger}_{B,2l}Y^C_{2l}Y^{\dagger}_{A,2l}Y^B_{2l}Y^{\dagger}_{C,2l} \nonumber \\
\nonumber \\
&-6\textrm{tr} Y^A_{2l-1}Y^{\dagger}_{B,2l-1}Y^B_{2l-1}Y^{\dagger}_{A,2l-1}Y^C_{2l-1}Y^{\dagger}_{C,2l-1}-6\textrm{tr} Y^A_{2l}Y^{\dagger}_{B,2l}Y^B_{2l}Y^{\dagger}_{A,2l}Y^C_{2l}Y^{\dagger}_{C,2l} \nonumber \\
&-6\textrm{tr} Y^A_{2l+1}Y^{\dagger}_{B,2l+1}Y^B_{2l+1}Y^{\dagger}_{A,2l+1}Y^C_{2l}Y^{\dagger}_{C,2l}-6\textrm{tr} Y^A_{2l}Y^{\dagger}_{B,2l}Y^B_{2l}Y^{\dagger}_{A,2l}Y^C_{2l+1}Y^{\dagger}_{C,2l+1} \nonumber \\
&-6\textrm{tr} Y^A_{2l-1}Y^{\dagger}_{B,2l}Y^B_{2l}Y^{\dagger}_{A,2l-1}Y^C_{2l-1}Y^{\dagger}_{C,2l-1}-6\textrm{tr} Y^A_{2l}Y^{\dagger}_{B,2l-1}Y^B_{2l-1}Y^{\dagger}_{A,2l}Y^C_{2l}Y^{\dagger}_{C,2l} \nonumber \\
&-6\textrm{tr} Y^A_{2l+1}Y^{\dagger}_{B,2l+2}Y^B_{2l+2}Y^{\dagger}_{A,2l+1}Y^C_{2l}Y^{\dagger}_{C,2l}-6\textrm{tr} Y^A_{2l}Y^{\dagger}_{B,2l-1}Y^B_{2l-1}Y^{\dagger}_{A,2l}Y^C_{2l+1}Y^{\dagger}_{C,2l+1} \Big],
\label{N4bos}
\end{align}
where we used $SU(2)$ doublets
\begin{align}
Y^A_l=\{ Z^{(l)} , W^{(l)\dagger} \} , \ \ Y^{\dagger}_{A,l}=\{ Z^{(l)\dagger} , W^{(l)} \}, \ \ \ \ (A=1,2)
\end{align}
for each link $l$. The quiver diagram of this theory is given in Figure \ref{N=4}. 

This theory has product gauge group $U(N)^{2n}$ and its moduli space is $Sym^N({\mathbb{C}}^4 / ({\mathbb{Z}}_{kn} \times {\mathbb{Z}}_n))$.
${\mathbb{Z}}_{nk}$ corresponds to the original ABJM orbifold action,
\begin{align}
y^1 \rightarrow e^{2\pi i/nk}y^1, \ \ y^2 \rightarrow e^{2\pi i/nk}y^2, \ \ y^3 \rightarrow e^{2\pi i /nk}y^3, \ \ y^4 \rightarrow e^{2\pi
 i/nk}y^4.
\end{align}
 Note that in order to have a correct moduli space, as explained in \cite{Terashima:2008ba}, the levels
 of the Chern-Simons terms in (\ref{N4QCS}) must be $\pm k$, not $\pm nk$. Another ${\mathbb{Z}}_n$ action is given by
\begin{align}
y^1 \rightarrow e^{2\pi i/n}y^1, \ \ y^2 \rightarrow y^2, \ \ y^3 \rightarrow e^{2\pi i /n}y^3, \ \ y^1 \rightarrow y^4.
\label{orb1}
\end{align}
This kind of further orbifolding is essential for deriving the Extended L-BLG theory from the ABJM theory. In \cite{Honma:2008jd}, we obtained a
 circle by taking a limit of the original ABJM orbifold action and rescaling the fields. Therefore, in a similar fashion, the emergence of an additional
 circle is expected in a suitable limit of ${\mathbb{Z}}_n$ action. Naively, it seems that the more we orbifold the ABJM theory, the more we have
 additional circles. However, in this paper, we only consider the case for one additional circle, namely, $T^2$ compactification of M-theory. We show that a
 proper scaling limit leads to the Extended L-BLG theory with $d=1$.
\section{Scaling limit of ${\cal{N}}=4$ quiver Chern-Simons theory}
\setcounter{equation}{0}
Here we explicitly show how the Extended L-BLG theory with $d=1$ is derived from a ${\cal{N}}=4$ quiver Chern-Simons theory (\ref{N4QCS}). 
First, we take linear combinations for the gauge fields as
\begin{align}
A^{(\pm)(2l-1)}_{\mu}=\frac{1}{2}(A^{(2l-1)}_{\mu} \pm A^{(2l+2s)}_{\mu}), \ \ \ \ (s \in {\mathbb{Z}})
\label{gaugeLC}
\end{align}
and decompose the bifundamental fields into trace and traceless parts as $Y=Y_0 {\bf{1}}_{N \times N}+\hat{Y}$. VEV $Y_0$ is
 interpreted as a classical position of the center of mass of the multiple M2-branes, and $\hat{Y}=\hat{Y}_a T^a$ is a fluctuation around it.
 $T^a$ is the generator of $SU(N)$. Next, we rescale the fields as
\begin{align}
&Y^1_{0 \ (2l-1)} \rightarrow \sqrt{\frac{n}{2}}Y^{(1)}_{0}, \ \ \ Y^1_{0 \ (2l)} \rightarrow \sqrt{\frac{n}{2}}Y^{(2)}_{0}, \ \ \ Y^2_{0 \
 (2l-1)} \rightarrow \sqrt{\frac{n}{2}}Y^{(3)}_{0}, \ \ \ Y^2_{0 \ (2l)} \rightarrow \sqrt{\frac{n}{2}}Y^{(4)}_{0}, \nonumber \\
&\hat{Y}^1_{ \ (2l-1)} \rightarrow \frac{q^{lm}}{\sqrt{n}}\frac{Y^{(1)}_{ \ (m)}}{\sqrt{2}}, \ \ \ \hat{Y}^1_{ \ (2l)} \rightarrow \frac{q^{lm}}{\sqrt
{n}}\frac{Y^{(2)}_{ \ (m)}}{\sqrt{2}}, \ \ \ \hat{Y}^2_{ \ (2l-1)} \rightarrow \frac{q^{lm}}{\sqrt{n}}\frac{Y^{(3)}_{ \ (m)}}{\sqrt{2}}, \ \ \ \hat
{Y}^2_{ \ (2l)} \rightarrow \frac{q^{lm}}{\sqrt{n}}\frac{Y^{(4)}_{ \ (m)}}{\sqrt{2}}, \nonumber \\
\nonumber \\
&A^{(+)(2l-1)}_{\mu} \rightarrow q^{lm}A_{\mu (m)}, \ \ \ A^{(-)(2l-1)}_{\mu} \rightarrow \frac{\pi}{n}q^{lm}A'_{\mu (m)}
\label{limit}
\end{align}
and finally take $n \rightarrow \infty$. Here, $q \equiv e^{\frac{2\pi i}{n}}$ and multiplying $q^{lm}$ corresponds to the Fourier transformation.
 The normalization is determined by $\sum_l q^{lm}=n\delta_{m,0}$.
Recalling that this ${\cal{N}}=4$ quiver CS theory describes multiple M2-branes at the singularity of an orbifold ${\mathbb{C}}^4 /
 ({\mathbb{Z}}_{nk} \times {\mathbb{Z}}_n)$, this scaling limit corresponds to locating the M2-branes far from the origin of the orbifold and
 simultaneously making each ${\mathbb{Z}}_{nk} , {\mathbb{Z}}_n$ identifications into the independent circle identifications. This is effctively the
 same as the ordinary $T^2$ compactification. Therefore, we can expect that the Extended L-BLG theory with $d=1$ emerges from this limit.

First, let us check the kinetic term. The covariant derivatives (\ref{coder1}) are scaled as
\begin{align}
&D_{\mu}Z_{(2l-1)} \rightarrow \frac{q^{lm}}{\sqrt{n}} \cdot \frac{1}{\sqrt{2}}\Big[ \partial_{\mu}Y^{(1)}_{(m)}+i[A_{\mu (n)},Y^{(1)}_{(m-n)}]-2\pi
 smA_{\mu (m)}Y^{(1)}_0+2\pi i A'_{\mu (m)}Y^{(1)}_0 +{\cal{O}}(n^{-1}) \Big] , \nonumber \\
&D_{\mu}Z_{(2l)} \rightarrow \frac{q^{lm}}{\sqrt{n}} \cdot \frac{1}{\sqrt{2}}\Big[ \partial_{\mu}Y^{(2)}_{(m)}+i[A_{\mu (n)},Y^{(2)}_{(m-n)}]-2\pi
 (s+1)m A_{\mu (m)}Y^{(2)}_0+2\pi i A'_{\mu (m)}Y^{(2)}_0 \nonumber \\
& \ \ \ \ \ \ \ \ \ \ \ \ \ \ \ \ \ \ \ \ \ \ \ \ \ \ \ \ \ \ +{\cal{O}}(n^{-1}) \Big] , \nonumber \\
&D_{\mu}W_{(2l-1)} \rightarrow \frac{1}{\sqrt{2}}\Big[ \frac{q^{-lm}}{\sqrt{n}}\partial_{\mu}Y^{(3) \dagger}_{(m)}+i\frac{q^{lm}}{\sqrt{n}}[A_{\mu
 (n)},Y^{(3) \dagger}_{(n-m)}]+\frac{2\pi sm}{\sqrt{n}}q^{lm}A_{\mu (m)}Y^{(3) \dagger}_0 \nonumber \\
& \ \ \ \ \ \ \ \ \ \ \ \ \ \ \ \ \ \ \ \ \ \ \ \ \ \ \ \ \ \ \ \ \ \ \ -i\frac{2\pi }{\sqrt{n}}q^{lm}A'_{\mu (m)}Y^{(3) \dagger}_0+{\cal{O}}(n^{-1}) \Big] ,
 \nonumber \\
&D_{\mu}W_{(2l)} \rightarrow \frac{1}{\sqrt{2}}\Big[ \frac{q^{-lm}}{\sqrt{n}}\partial_{\mu}Y^{(4) \dagger}_{(m)}+i\frac{q^{lm}}{\sqrt{n}}[A_{\mu
 (n)},Y^{(4) \dagger}_{(n-m)}]+\frac{2\pi (s+1)m}{\sqrt{n}} q^{lm}A_{\mu (m)}Y^{(4) \dagger}_0 \nonumber \\
& \ \ \ \ \ \ \ \ \ \ \ \ \ \ \ \ \ \ \ \ \ \ \ \ \ \ \ \ \ \ \ \ -i\frac{2\pi}{\sqrt{n}} q^{lm}A'_{\mu (m)}Y^{(4) \dagger}_0+{\cal{O}}(n^{-1}) \Big].
\label{scaco1}
\end{align}
The ${\cal{O}}(n^{-1})$ terms do not contribute to the action in the limit $n \rightarrow \infty$.

In our notation, complex scalar fields are decomposed to real fields as
\begin{align}
&Y^{(A)}_0=X^A_0+iX^{A+4}_0, \nonumber \\
&Y^{(A)}_{(m)}=i\hat{X}^A_{(m)}-\hat{X}^{A+4}_{(m)}.
\label{dec}
\end{align}
We note that hermitian conjugation changes the sign of the label $m$ such as 
\begin{align}
Y^{(A) \dagger}_{(m)}=-i\hat{X}^A_{(-m)}-\hat{X}^{A+4}_{(-m)} \ , \ \ \ \ \ A^{\dagger}_{\mu (m)}=A_{\mu (-m)}.
\label{her}
\end{align}
Combining (\ref{scaco1}),(\ref{dec}) and (\ref{her}), we can write out a rescaled kinetic term using real fields.
 Let us compare this kinetic term with that of the Extended L-BLG theory given by
\begin{align}
-\frac{1}{2}(D_{\mu}X^I_{(-m)})&(D^{\mu}X^I_{(m)}) = -\frac{1}{2}\partial_{\mu}X^I_{(-m)}\partial^{\mu}X^I_{(m)}-i\partial_{\mu}X^I_{(-m)}
[A^{\mu}_{(n)},X^I_{(m-n)}] \nonumber \\
&-\frac{1}{2}[X^I_{(-m+n)},A_{\mu (-n)}][A^{\mu}_{(k)},X^I_{(m-k)}]+A'^{\mu}_{(m)}\lambda^{I0}\partial_{\mu}X^I_{(-m)}
+imA^{\mu}_{(m)}\lambda^{I1}\partial_{\mu}X^I_{(-m)} \nonumber \\
&-iA'^{\mu}_{(m)}\lambda^{I0}[X^1_{(-m+n)},A_{\mu (-n)}]+mA^{\mu}_{(m)}\lambda^{I1}[X^I_{(-m+n)},A_{\mu (-n)}] \nonumber \\
&-\frac{1}{2}A'_{\mu (-m)}A'^{\mu}_{(m)}(\lambda^{I0})^2-\frac{1}{2}m^2A_{\mu (-m)}A^{\mu}_{(m)}(\lambda^{I1})^2
+imA_{\mu (-m)}A'^{\mu}_{(m)}\lambda^{I0}\lambda^{I1}.
\label{ELkine}
\end{align}
Then, we see that if we identify 
\begin{align}
&\lambda^{I0}=-2\pi ( X^1_0, X^2_0, X^3_0, X^4_0, X^5_0, X^6_0, X^7_0, X^8_0 ), \nonumber \\
&\lambda^{I1}=-2\pi \Big( sX^1_0, (s+1)X^2_0, sX^3_0, (s+1)X^4_0, sX^5_0, (s+1)X^6_0, sX^7_0, (s+1)X^8_0 \Big),
\label{VEV1}
\end{align}
both kinetic terms completely agree.

For the Chern-Simons term, we can show the agreement easily:
\begin{align}
&\frac{k}{4\pi} \epsilon^{\mu\nu\lambda} \Big[ A^{(2l-1)}_{\mu}\partial_{\nu}A^{(2l-1)}_{\lambda}+\frac{2i}{3}A^{(2l-1)}_{\mu}A^{(2l-1)}_{\nu}A^{(2l-1)}_{\lambda} -A^{(2l)}_{\mu}\partial_{\nu}A^{(2l)}_{\lambda}-\frac{2i}{3}A^{(2l)}_{\mu}A^{(2l)}_{\nu}A^{(2l)}_{\lambda} \Big] \nonumber \\
&=\frac{k}{2\pi}\epsilon^{\mu\nu\lambda}A^{(-)(2l-1)}_{\mu}F^{(2l-1)}_{\nu\lambda}+\frac{4i}{3}\epsilon^{\mu\nu\lambda}A^{(-)(2l-1)}_
{\mu}A^{(-)(2l-1)}_{\nu}A^{(-)(2l-1)}_{\lambda} \nonumber \\
&=\frac{k}{2} \epsilon^{\mu\nu\lambda}\frac{q^{l(m+n)}}{n}A'_{\mu (m)}F_{\nu\lambda (n)}+\frac{ik}{3\pi}\epsilon^{\mu\nu\lambda}\frac{q^{lm}}{n^3}
A'_{\mu (n)}A'_{\nu (k)}A'_{\lambda (m-n-k)} \nonumber \\
&\rightarrow \frac{k}{2}\epsilon^{\mu\nu\lambda}A'_{\mu (m)}F_{\nu\lambda (-m)},
\label{CS}
\end{align}
where $F_{\nu\lambda}^{(2l-1)}=\partial_{\nu}A_{\lambda}^{(+)(2l-1)}-\partial_{\lambda}A_{\nu}^{(+)(2l-1)}+i[ A_{\nu}^{(+)(2l-1)},
 A_{\lambda}^{(+)(2l-1)}]$.
 Note that we have chosen $k=1$ in the BLG side.

In the Extended L-BLG theory, VEVs $\lambda^{IA}$ are related to the metric of two-torus as (\ref{met}).
 By constructing the metric $G^{AB}$ from (\ref{VEV1}), we see that the metric components are connected as
\begin{align}
G^{11}=-s(s+1)G^{00}+(2s+1)G^{01}.
\label{cons1}
\end{align}
Thus, in the scaling limit of the ${\cal{N}}=4$ quiver CS theory, only a specific class of the $T^2$ compactification is realizable.
 This is because we have chosen a particular ${\mathbb{Z}}_n$ orbifold. Owing to the constraint (\ref{cons1}), the complexified coupling constant
 $\tau$ of the resultant D3-brane theory is limited to the one that depends on only one real variable. We will return to this point in Section 5.

Now, let us check the potential term. By decomposing the matter fields $Y^A_{l}$ into the trace part $Y^A_{0}$ and the traceless part $\hat{Y}^A_l$,
 the bosonic sextic potential term becomes $V_{bos} = \sum^6_{s=0} V^{(s)}_{bos}$, where $V^{(s)}_{bos}$ contains $s$ $Y_0$ fields and $(6-s)$
 $\hat{Y}$ fields. It can be easily checked that $V^{(6)}_{bos}$ and $V^{(5)}_{bos}$ are indentically zero.
 Since $V^{(s)}_{bos}$ scales as $n^{\frac{s}{2}-\frac{6-s}{2}+1}=n^{s-2}$ in our limit (\ref{limit}), $V^{(0)}_{bos}$ and $V^{(1)}_{bos}$ vanish.
 Note that there is an additional factor $n$ that comes from the relation $\sum_l q^{lm} = n\delta_{m,0}$. Therefore, the remaining terms are
 $V^{(2)}_{bos}$, $V^{(3)}_{bos}$, and $V^{(4)}_{bos}$.

 First, we consider the scaling limit of $V^{(2)}_{bos}$. In this case, we can utilize the result in \cite{Honma:2008jd} and obtain the scaling limit easily.
 The key point is the fact that the relative difference of label $l$ becomes ${\cal{O}}(n^{-\frac{3}{2}})$ under the expansion $q^{lm} = 1 +
 \frac{2\pi i lm}{n} + {\cal{O}}({n^{-2}})$:
 \begin{eqnarray}
  (\hat{Y}_{2l}-\hat{Y}_{2(l+k)}) \rightarrow
   \frac{q^{lm}}{\sqrt{n}}(Y_m - q^{km} Y_m) =
   {\cal{O}}(n^{-\frac{3}{2}}).
 \end{eqnarray}
 This means that in the scaling limit of $V^{(2)}_{bos}$, the relative difference between the labels of $\hat{Y}_{2l} \ (\textrm{or} \ \hat{Y}_{2l-1}
 \ \textrm{in the odd case})$
 does not contribute to the result. To show this explicitly, let us consider the scaling limit of the following substraction:
\begin{align}
 Y_{0,2l}Y^{\dagger}_{0 \ 2l} \hat{Y}_{2l} \hat{Y}^{\dagger}_{2l}
  (\hat{Y}_{2(l+k)} - \hat{Y}_{2l}) \hat{Y}^{\dagger}_{2l} \rightarrow {\cal{O}}(n^{-1}) = 0.
\label{subt}
\end{align}
Note that if the numbers of $Y_{0,l}$ and $\hat{Y}_l$ are different, the situation entirely changes. Indeed, for the scaling limit of $V^{(3)}_{bos}$
 and $V^{(4)}_{bos}$, the relative difference between the labels of $\hat{Y}_l$ is essential.
 The relation like (\ref{subt}) holds in all the terms of (\ref{N4bos}). Therefore, even if we replace all the $Y_{2(l+k)-1}^A$ with $Y^A_{2l-1}$
 (and $Y^A_{2(l+k)}$ with $Y^A_{2l}$) in (\ref{N4bos}), the resultant potential gives the same scaling limit as long as we focus on the 
$Y_{0,l}$-squared term. We denote this new potential as $V'$
\begin{align}
 V' = &-\frac{4 {\pi}^2}{3 k^2}
 \Big[ \ 
\textrm{tr} Y^A_{2l}Y^{\dagger}_{A,2l}Y^B_{2l}Y^{\dagger}_{B,2l}Y^C_{2l}Y^{\dagger}_{C,2l}+3\textrm{tr} Y^A_{2l}Y^{\dagger}_{A,2l}Y^B_{2l}Y^{\dagger}_{B,2l}Y^C_{2l-1}Y^{\dagger}_{C,2l-1} \nonumber \\
&+3\textrm{tr} Y^A_{2l}Y^{\dagger}_{A,2l}Y^B_{2l-1}Y^{\dagger}_{B,2l-1}Y^C_{2l-1}Y^{\dagger}_{C,2l-1}+\textrm{tr} Y^A_{2l-1}Y^{\dagger}_{A,2l-1}Y^B_{2l-1}Y^{\dagger}_{B,2l-1}Y^C_{2l-1}Y^{\dagger}_{C,2l-1} \nonumber \\
\nonumber \\
&+\textrm{tr} Y^{\dagger}_{A,2l-1}Y^A_{2l-1}Y^{\dagger}_{B,2l-1}Y^B_{2l-1}Y^{\dagger}_{C,2l-1}Y^C_{2l-1}+3\textrm{tr} Y^{\dagger}_{A,2l-1}Y^A_{2l-1}Y^{\dagger}_{B,2l-1}Y^B_{2l-1}Y^{\dagger}_{C,2l}Y^C_{2l} \nonumber \\
&+3\textrm{tr} Y^{\dagger}_{A,2l-1}Y^A_{2l-1}Y^{\dagger}_{B,2l}Y^B_{2l}Y^{\dagger}_{C,2l}Y^C_{2l}+\textrm{tr} Y^{\dagger}_{A,2l}Y^A_{2l}Y^{\dagger}_{B,2l}Y^B_{2l}Y^{\dagger}_{C,2l}Y^C_{2l} \nonumber \\
\nonumber \\
&+4\textrm{tr} Y^A_{2l-1}Y^{\dagger}_{B,2l-1}Y^C_{2l-1}Y^{\dagger}_{A,2l-1}Y^B_{2l-1}Y^{\dagger}_{C,2l-1}+12\textrm{tr} Y^A_{2l}Y^{\dagger}_{B,2l}Y^C_{2l-1}Y^{\dagger}_{A,2l}Y^B_{2l}Y^{\dagger}_{C,2l-1} \nonumber \\
&+12\textrm{tr} Y^A_{2l-1}Y^{\dagger}_{B,2l-1}Y^C_{2l}Y^{\dagger}_{A,2l-1}Y^B_{2l-1}Y^{\dagger}_{C,2l}+4\textrm{tr} Y^A_{2l}Y^{\dagger}_{B,2l}Y^C_{2l}Y^{\dagger}_{A,2l}Y^B_{2l}Y^{\dagger}_{C,2l} \nonumber \\
\nonumber \\
&-6\textrm{tr} Y^A_{2l-1}Y^{\dagger}_{B,2l-1}Y^B_{2l-1}Y^{\dagger}_{A,2l-1}Y^C_{2l-1}Y^{\dagger}_{C,2l-1}-6\textrm{tr} Y^A_{2l}Y^{\dagger}_{B,2l}Y^B_{2l}Y^{\dagger}_{A,2l}Y^C_{2l}Y^{\dagger}_{C,2l} \nonumber \\
&-6\textrm{tr} Y^A_{2l-1}Y^{\dagger}_{B,2l-1}Y^B_{2l-1}Y^{\dagger}_{A,2l-1}Y^C_{2l}Y^{\dagger}_{C,2l}-6\textrm{tr} Y^A_{2l}Y^{\dagger}_{B,2l}Y^B_{2l}Y^{\dagger}_{A,2l}Y^C_{2l-1}Y^{\dagger}_{C,2l-1} \nonumber \\
&-6\textrm{tr} Y^A_{2l-1}Y^{\dagger}_{B,2l}Y^B_{2l}Y^{\dagger}_{A,2l-1}Y^C_{2l-1}Y^{\dagger}_{C,2l-1}-6\textrm{tr} Y^A_{2l}Y^{\dagger}_{B,2l-1}Y^B_{2l-1}Y^{\dagger}_{A,2l}Y^C_{2l}Y^{\dagger}_{C,2l} \nonumber \\
&-6\textrm{tr} Y^A_{2l-1}Y^{\dagger}_{B,2l}Y^B_{2l}Y^{\dagger}_{A,2l-1}Y^C_{2l}Y^{\dagger}_{C,2l}-6\textrm{tr} Y^A_{2l}Y^{\dagger}_{B,2l-1}Y^B_{2l-1}Y^{\dagger}_{A,2l}Y^C_{2l-1}Y^{\dagger}_{C,2l-1} \Big].
\end{align}

$V'$ is convenient because it can be simplified.
If we rewrite each field as 
 \begin{eqnarray}
  Y^1_{2l-1} \rightarrow Y^1_l, \ \ \ Y^1_{2l} \rightarrow Y^2_l,
   \ \ \ Y^2_{2l-1} \rightarrow Y^3_l, \ \ \ Y^2_{2l} \rightarrow Y^4_l,
 \end{eqnarray}
$V'$ becomes
\begin{eqnarray}
 & & -\frac{4\pi^2}{3k^2} \Big[
  Y^{A^{\prime}}_l Y^{\dagger}_{A^{\prime},l} Y^{B^{\prime}}_l Y_{B^{\prime},l}^{\dagger} Y^{C^{\prime}}_l Y^{\dagger}_{C^{\prime},l}
  + Y^{\dagger}_{A^{\prime},l} Y^{A^{\prime}}_l Y^{\dagger}_{B^{\prime},l} Y^{B^{\prime}}_l Y_{C^{\prime},l}^{\dagger}
  Y^{C^{\prime}}_l \nonumber \\ 
  & & \ \ \ \ \ \ \ \ \ \ \ \ \ 
   + 4 Y^{A^{\prime}}_l Y^{\dagger}_{B^{\prime},l} Y^{C^{\prime}}_l Y^{\dagger}_{A^{\prime},l} Y^{B^{\prime}}_l
   Y_{C^{\prime},l}^{\dagger} 
   - 6 Y^{A^{\prime}}_l Y^{\dagger}_{B^{\prime},l} Y^{B^{\prime}}_l Y^{\dagger}_{A^{\prime},l} Y^{C^{\prime}}_l
   Y_{C^{\prime},l}^{\dagger}\Big],
\end{eqnarray}
 where $A^{\prime},B^{\prime},C^{\prime}=1,\cdots,4$. This is just the original ABJM potential with an extra label $l$. The scaling limit of the
 original ABJM bosonic potential is already obtained in \cite{Honma:2008jd} and the result is
\begin{eqnarray}
  \textrm{tr} (X^I_0)^2 ([P^{IK}X^K,P^{JL}X^L])^2.
\label{scaABJMbos}
\end{eqnarray}
Using this result, we can obtain the scaling limit of $V_{bos}^{(2)}$:
\begin{eqnarray}
 V^{(2)}_{bos} \rightarrow -\frac{\pi^2}{k^2}(X^I_0)^2 [P^{IK}X^K_{(m)}, P^{JL}X^L_{(-m)}].  
\end{eqnarray}
This agrees with the last term of (\ref{ELbos}).

Next we consider the scaling limit of $V^{(4)}_{bos}$ and $V^{(3)}_{bos}$.
 As before, we can decompose $V'$ as
$V^\prime = \sum^{6}_{s=0} V^{\prime(s)}$.
Using the same argument, we see that only
$V^{\prime(2)}$, $V^{\prime(3)}$, and $V^{\prime(4)}$ remain in
the scaling limit.

 In (\ref{scaABJMbos}), more insertion of $X^K_0$ to $X^K$ gives zero.
 Therefore, $V^{\prime(3)}$ and $V^{\prime(4)}$ are zero.
 This means that the scaling limit of $V_{bos}-V^\prime$ is the same as the
 scaling limit of $V^{(3)}_{bos}+V^{(4)}_{bos}$. It is convenient to consider
 $V_{bos}-V_0$ because it is much simpler than $V_{bos}$ itself.
 The explicit form of $V_{bos}-V^\prime$ is given by
\begin{eqnarray}
& V_{bos}-V^\prime = V_1 + V_2,
 \end{eqnarray}
 where
 \begin{align}
V_1 = -\frac{4\pi^2}{3k^2}  \textrm{tr}& \big[
  3 Y^A_{2l-1}Y^\dagger_{A,2l-1}Y^B_{2l-1}Y^\dagger_{B,2l-1}
  (Y^C_{2l-2}Y^\dagger_{C,2l-2} - Y^C_{2l} Y^\dagger_{C,2l})
  \ \ \ \ \nonumber \\
& +12 Y^C_{2l} Y^\dagger_{A,2l-1} Y^B_{2l-1} Y^\dagger_{C,2l}
  ( Y^A_{2l+1}Y^\dagger_{B,2l+1} - Y^A_{2l-1}Y^\dagger_{B,2l-1})
  \nonumber \\
& - 6 Y^A_{2l-1} Y^\dagger_{B,2l-1} Y^B_{2l-1} Y^\dagger_{A,2l-1}
  (Y^C_{2l-2}Y^\dagger_{C,2l-2} - Y^C_{2l}Y^\dagger_{C,2l} )
  \ \nonumber \\
&  - 6 Y^C_{2l}Y^\dagger_{A,2l-1}Y^A_{2l-1}Y^\dagger_{C,2l}
  (Y^B_{2l+1}Y^\dagger_{B,2l+1} - Y^B_{2l-1}Y^\dagger_{B,2l-1}) \big],
\end{align}
and
\begin{align}
V_2 = -\frac{4\pi^2}{3k^2} \textrm{tr}& \big[
   3 Y^A_{2l}Y^\dagger_{A,2l}Y^B_{2l}Y^\dagger_{B,2l}
  (Y^C_{2l+1}Y^\dagger_{C,2l+1} - Y^C_{2l-1} Y^\dagger_{C,2l-1})
  \ \ \ \ \nonumber \\
& +12 Y^C_{2l-1} Y^\dagger_{A,2l} Y^B_{2l} Y^\dagger_{C,2l-1}
  ( Y^A_{2l-2}Y^\dagger_{B,2l-2} - Y^A_{2l}Y^\dagger_{B,2l})
  \nonumber \\
& - 6 Y^A_{2l} Y^\dagger_{B,2l} Y^B_{2l} Y^\dagger_{A,2l}
  (Y^C_{2l+1}Y^\dagger_{C,2l+1} - Y^C_{2l-1}Y^\dagger_{C,2l-1} )
  \ \nonumber \\
&  - 6 Y^C_{2l-1}Y^\dagger_{A,2l}Y^A_{2l}Y^\dagger_{C,2l-1}
  (Y^B_{2l-2}Y^\dagger_{B,2l-2} - Y^B_{2l}Y^\dagger_{B,2l}) \big].\ \ 
\end{align}
Note that $V_1$ and $V_2$ can be translated into each other by exchanging
$Y^A_{2l}$ for $Y^A_{2l-1}$ and $Y^A_{2l-2}$ for $Y^A_{2l+1}$.
Since the rescaling rule (\ref{limit}) is written as
\begin{eqnarray}
 Y^A_{2l} \rightarrow
  \frac{q^{lm}}{\sqrt{n}}\frac{Y^{2A}_{m}}{\sqrt{2}}, \ \ 
  Y^A_{2l-1} \rightarrow
  \frac{q^{lm}}{\sqrt{n}}\frac{Y^{2A-1}_{m}}{\sqrt{2}}, \ \
  Y^A_{2l-2} \rightarrow q^{-m}
  \frac{q^{lm}}{\sqrt{n}}\frac{Y^{2A}_{m}}{\sqrt{2}}, \ \
  Y^A_{2l+1} \rightarrow q^m
  \frac{q^{lm}}{\sqrt{n}}\frac{Y^{2A-1}_{m}}{\sqrt{2}},
\label{mato}
\end{eqnarray}
the above translation corresponds to a
translation between $Y^{2A}_m$ and $Y^{2A+1}_m$.

Therefore, to obtain the scaling limit of $V_1$ and $V_2$, we only need to calculate one of them.
 The other one is obtained from the translation.
 
With the above simplifications, the scaling limit of $V_{bos}^{(4)}$ can be calculated more easily. The result is
\begin{eqnarray}
 \frac{m^2 (16\pi^4)}{2} 
 (X_{0}^{2C}X_0^{2C}X_0^{2A-1}X_0^{2A-1}
 \hat{X}^{2B-1}_{(i,m)}\hat{X}^{2B-1}_{(i,-m)} 
 - X^{2C}_0 X^{2C}_0 X^{2A-1}_0 X^{2B-1}_0
 \hat{X}^{2A-1}_{(i,m)}\hat{X}^{2B-1}_{(i,-m)} \nonumber \\
 + X_{0}^{2C-1}X_0^{2C-1}X_0^{2A}X_0^{2A}
 \hat{X}^{2B}_{(i,m)}\hat{X}^{2B}_{(i,-m)} 
 - X^{2C-1}_0 X^{2C-1}_0 X^{2A}_0 X^{2B}_0
 \hat{X}^{2A}_{(i,m)}\hat{X}^{2B}_{(i,-m)}).
 \label{scaV4}
\end{eqnarray}
This is just the first term of (\ref{ELbos}) with the assignment (\ref{VEV1}).
To see how the above terms come from the Extended L-BLG
potential, it is convenient to use an expression
\begin{eqnarray}
 m^2 \lambda^{[I0} \lambda^{J1} X^{K]}_{i,m}
  \lambda^{[I0} \lambda^{J1} X^{K]}_{i,-m}
\end{eqnarray}
and substitute (\ref{VEV1}) into this term. Then, we obtain (\ref{scaV4}).
 Note that the result does not depend on $s$, because the $s$-dependent part of $\lambda^{I1}$ is proportional to $\lambda^{I0}$
 and the indices $I,J,$ and $K$ are antisymmetrized so that $s$ dependent terms are cancelled.

Similarly, the scaling limit of $V^{(3)}_{bos}$ is given by
 \begin{eqnarray}
&&  (2\pi)^3 \textrm{tr} \big\{ 
(2m+n) X^{2C}_0 X^{2A-1}_0 X^{2B-1}_0
   \hat{X}^{2A-1}_m [\hat{X}^{2C}_n,\hat{X}^{2B-1}_{-m-n}]
   \ \ \ \ \ \ \ \ \ \nonumber \\
&& \ \ \ \ \ \ \ \ \ \   + m X^{2C}_0 X^{2C}_0 X^{2B-1}_0 
   \hat{X}^{2A-1}_{m}[\hat{X}^{2B-1}_n,\hat{X}^{2A-1}_{-m-n}] -
   mX^{2C}_0 X^{2B-1}_0 X^{2B-1}_0
   \hat{X}^{2A-1}_m [\hat{X}^{2C}_n,\hat{X}^{2A-1}_{-m-n}] 
      \nonumber \\
&& \ \ \ \ \ \ \ \ \ \ - (2m+n) X^{2C-1}_0 X^{2A}_0 X^{2B}_0
   \hat{X}^{2A}_m [\hat{X}^{2C-1}_n,\hat{X}^{2B}_{-m-n}]
\nonumber \\ 
&& \ \ \ \ \ \ \ \ \ \ - m
   X^{2C-1}_0 X^{2C-1}_0 X^{2B}_0 
   \hat{X}^{2A}_{m}[\hat{X}^{2B}_n,\hat{X}^{2A}_{-m-n}]
   + mX^{2C-1}_0 X^{2B}_0 X^{2B}_0
   \hat{X}^{2A}_m [\hat{X}^{2C-1}_n,\hat{X}^{2A}_{-m-n}] 
 \big\}. \nonumber \\
 \label{scaV3}
 \end{eqnarray}
Note that the overall signs of $V^{(3)}_1$ and $V^{(3)}_2$ are opposite owing to the factors $q^{\pm m}$ in (\ref{mato}).
 (\ref{scaV3}) agrees with the second term of (\ref{ELbos}). 

\paragraph{Fermionic sector}
We have seen the agreement of the bosonic sector. Here, we consider the fermionic sector of the ${\cal{N}}=4$ quiver CS theory and confirm the
 emergence of the Extended L-BLG theory. The nontrivial part is the fermionic potential.

In the Extended L-BLG theory, the fermionic interaction term is given by
\begin{align}
L_{int}&=\frac{m_a}{4}\bar{\psi}_{(i\vec{-m})}( \Gamma_{IJ}\lambda^{I0}\lambda^{Ja})\psi_{(i,\vec{m})}+\frac{1}{4}\bar{\psi}_{(i\vec{m})}
\lambda^{I0} [X^J , \Gamma_{IJ}\psi]_{(i,-\vec{m})}.
\label{ELfer}
\end{align}
Substituting (\ref{Xdec}) into (\ref{ELfer}), we can indeed obtain the fermionic sector of the Dp-brane action.

On the other hand, the fermionic potential of the ${\cal{N}}=4$ quiver CS theory is given by
\begin{align}
V_{ferm}=-\frac{iL}{4}\textrm{tr} \Big[ &Y^{\dagger}_{A , 2l-1}Y^A_{2l-1}\Psi^{B \dagger}_{2l-1}\Psi_{B ,2l-1}+
Y^{\dagger}_{A , 2l-1}Y^A_{2l-1}\Psi^{B \dagger}_{2l}\Psi_{B ,2l} \nonumber \\
&+Y^{\dagger}_{A , 2l}Y^A_{2l}\Psi^{B \dagger}_{2l-1}\Psi_{B ,2l-1}+
Y^{\dagger}_{A , 2l}Y^A_{2l}\Psi^{B \dagger}_{2l}\Psi_{B ,2l} \nonumber \\
\nonumber \\
&-Y^A_{2l-1}Y^{\dagger}_{A,2l-1}\Psi_{B,2l-1}\Psi^{B \dagger}_{2l-1}
-Y^A_{2l+1}Y^{\dagger}_{A,2l+1}\Psi_{B,2l}\Psi^{B \dagger}_{2l} \nonumber \\
&-Y^A_{2l}Y^{\dagger}_{A,2l}\Psi_{B,2l+1}\Psi^{B \dagger}_{2l+1}
-Y^A_{2l}Y^{\dagger}_{A,2l}\Psi_{B,2l}\Psi^{B \dagger}_{2l} \nonumber \\
\nonumber \\
&+2Y^A_{2l-1}Y^{\dagger}_{B,2l}\Psi_{A,2l}\Psi^{B \dagger}_{2l-1}
+2Y^A_{2l}Y^{\dagger}_{B,2l-1}\Psi_{A,2l-1}\Psi^{B \dagger}_{2l} \nonumber \\
&+2Y^A_{2l}Y^{\dagger}_{B,2l}\Psi_{A,2l+1}\Psi^{B \dagger}_{2l+1}
+2Y^A_{2l+1}Y^{\dagger}_{B,2l+1}\Psi_{A,2l}\Psi^{B \dagger}_{2l} \nonumber \\
\nonumber \\
&-2Y^{\dagger}_{A,2l-1}Y^B_{2l-1}\Psi^{A \dagger}_{2l}\Psi_{B,2l}
-2Y^{\dagger}_{A,2l}Y^B_{2l}\Psi^{A \dagger}_{2l-1}\Psi_{B,2l-1} \nonumber \\
&-2Y^{\dagger}_{A,2l}Y^B_{2l+1}\Psi^{A \dagger}_{2l+1}\Psi_{B,2l}
-2Y^{\dagger}_{A,2l+1}Y^B_{2l}\Psi^{A \dagger}_{2l}\Psi_{B,2l+1} \nonumber \\
\nonumber \\
&-\epsilon^{AB}\epsilon^{CD}Y^{\dagger}_{A,2l-1}\Psi_{C,2l-1}Y^{\dagger}_{B,2l-1}\Psi_{D,2l-1}
-\epsilon^{AB}\epsilon^{CD}Y^{\dagger}_{A,2l}\Psi_{C,2l}Y^{\dagger}_{B,2l}\Psi_{D,2l} \nonumber \\
&+2\epsilon^{AB}\epsilon^{CD}Y^{\dagger}_{A,2l-1}\Psi_{C,2l-1}Y^{\dagger}_{D,2l}\Psi_{B,2l}
+2\epsilon^{AB}\epsilon^{CD}Y^{\dagger}_{A,2l+1}\Psi_{B,2l}Y^{\dagger}_{C,2l}\Psi_{D,2l+1} \nonumber \\
\nonumber \\
&+\epsilon_{AB}\epsilon_{CD}Y^A_{2l-1}\Psi^{C \dagger}_{2l-1}Y^B_{2l-1}\Psi^{D \dagger}_{2l-1}
+\epsilon_{AB}\epsilon_{CD}Y^A_{2l}\Psi^{C \dagger}_{2l}Y^B_{2l}\Psi^{D \dagger}_{2l} \nonumber \\
&-2\epsilon_{AB}\epsilon_{CD}Y^A_{2l-1}\Psi^{B \dagger}_{2l}Y^C_{2l}\Psi^{D \dagger}_{2l-1}
-2\epsilon_{AB}\epsilon_{CD}Y^A_{2l+1}\Psi^{C \dagger}_{2l+1}Y^D_{2l}\Psi^{B \dagger}_{2l} \Big],
\label{N4fer}
\end{align}
where $\epsilon^{12}=-\epsilon_{12}=1$ and we used doublets
\begin{align}
Y^A_l=\{ Z^{(l)} , W^{(l) \dagger} \} \ \ , \ \ \Psi_{A,l}=\{ (-1)^{l-1}e^{-i\pi /4}\zeta^{(l)} , (-1)^l e^{i\pi /4} \omega^{(l) \dagger} \}. \ \
 \ \ \ (A=1,2)
\end{align}
The label $l$ of $\zeta^{(l)}$ and $\omega^{(l)}$ was determined from the following orbifold projection of the $nN \times nN$ ABJM fermions:
\begin{align}
&\zeta^1 = \begin{pmatrix}
                 0 & \zeta^{(1)} & & & \\
                    & 0 & \zeta^{(3)} & & \\
                    & & 0 & \ddots & \\
                    & & & 0 & \zeta^{(2n-3)} \\
                    \zeta^{(2n-1)} & & & & 0
               \end{pmatrix},
\ \ \ \omega_1 = \begin{pmatrix}
                 0 & & & & \omega^{(2n-1)} \\
                  \omega^{(1)}  & 0 & & & \\
                    & \omega^{(3)} & 0 & & \\
                    & & \ddots & 0 & \\
                    & & & \omega^{(2n-3)} & 0
               \end{pmatrix}, \nonumber \\
\nonumber \\
&\zeta^2=\textrm{diag}(\zeta^{(2n)},\zeta^{(2)}, \cdots , \zeta^{(2n-2)}) \ ,
 \ \ \ \ \ \ \omega_2=\textrm{diag}(\omega^{(2n)},\omega^{(2)}, \cdots , \omega^{(2n-2)}).
\end{align}
Each $\zeta^{(l)}$ and $\omega^{(l)} \ (l=1,2, \cdots, 2n)$ are $N \times N$ matrices.

Now, we investigate the scaling limit of (\ref{N4fer}). The appropriate rescalings of the fermions are given by
\begin{align}
\Psi^1_{(2l-1)} \rightarrow \frac{q^{lm}}{\sqrt{n}}\frac{\Psi^{(2)}_{(m)}}{2} \ ,
 \ \Psi^1_{(2l)} \rightarrow \frac{q^{lm}}{\sqrt{n}}\frac{\Psi^{(1)}_{(m)}}{2} \ ,
 \ \Psi^2_{(2l-1)} \rightarrow \frac{q^{(l-2)m}}{\sqrt{n}}\frac{\Psi^{(4)}_{(m)}}{2} \ ,
 \ \Psi^2_{(2l)} \rightarrow \frac{q^{lm}}{\sqrt{n}}\frac{\Psi^{(3)}_{(m)}}{2}.
\end{align}
In analogy with the bosonic potential, after the decomposition $Y^A_{(l)}=Y^{A}_0 {\bf{1}}_{N \times N}+\hat{Y}^A_{(l)}$, the fermionic potential
 becomes
 $V_{ferm} = \sum^2_{s=0} V^{(s)}_{ferm}$, where $V^{(s)}_{ferm}$ contains $s$ $Y_0$ fields and $(2-s)$ $\hat{Y}$ fields. Obviously,
 $V^{(0)}_{ferm}$ vanishes in the limit $n \rightarrow \infty$. Thus, the remaining terms are $V^{(1)}_{ferm}$ and $V^{(2)}_{ferm}$.

First, let us consider the $V^{(2)}_{ferm}$ term. For simplicity, we consider the case where only the $Y^{(1)}_0$ and $Y^{(2)}_0$ are nonzero.
 Then the surviving terms in the limit $n \rightarrow \infty$ are summarized as
\begin{align}
\frac{4\pi^2 m}{k} \textrm{tr} \Big[ &2Y^{(2) \dagger}_0Y^{(1)}_0\Psi^{(2) \dagger}_{(m)}\Psi^{(1)}_{(m)}-
2Y^{(1) \dagger}_0Y^{(2)}_0\Psi^{(1) \dagger}_{(m)}\Psi^{(2)}_{(m)} \nonumber \\
&-2Y^{(1) \dagger}_0Y^{(2) \dagger}_0\Psi^{(3)}_{(-m)}\Psi^{(4)}_{(m)}+2Y^{(1)}_0Y^{(2)}_0\Psi^{(4) \dagger}_{(m)}\Psi^{(3) \dagger}_{(-m)}
 \Big].
\end{align}
After the decomposition of the fermions into the 2-component Majorana spinors as
\begin{align}
\Psi_{A (m)}=i\chi_{A (m)}-\chi_{A+4 (m)},
\end{align}
we obtain various bilinear terms of $\chi_{1 (m)} , \cdots, \chi_{8 (m)}$. Using the appropriate Gamma matrices, the assignment (\ref{VEV1}), and the identification
 $\psi^T_{(m)}=(\chi^T_{1(m)}, \cdots , \chi^T_{8(m)})$, we can show that these bilinear terms agree with the first term of (\ref{ELfer}). The explicit forms of
 the Gamma matrices are written in the Appendix.
 
As for the $V^{(1)}_{ferm}$ term, the situation is the same as the $V^{(2)}_{bos}$ term. In the scaling limit, we just need to
consider whether the index $l$ of $Y_l^A$ and $\Psi_l^A$ is odd or even, namely, we can replace all the $Y^A_{l'} \ (l' \in
 {\mathbb{Z}})$ with $Y^A_{2l-1}$ or $Y^A_{2l}$. This denotes that the fermion potential of the original ABJM theory with the additional
 labels $l$
\begin{align}
&-\frac{2\pi i}{k} \tr \, [Y^{\dagger}_{A,l} Y^A_l
 \Psi^{B\dagger}_l \Psi_{B,l} - Y^A_l Y^{\dagger}_{A,l} \Psi_{B,l} \Psi^{B\dagger}_l + 2
 Y^A_l Y^{\dagger}_{B,l} \Psi_{A,l} \psi^{B\dagger}_l - 2 Y^{\dagger}_{A,l} Y^B_l
 \Psi^{A\dagger}_l \Psi_{B,l} \nonumber \\
&  \ \ \ \ \ \ \ \ \ \ \ \ \ \ \ \ 
+ \epsilon^{ABCD} Y^{\dagger}_{A,l} \Psi_{B,l} Y^{\dagger}_{C,l} \Psi_{D,l} -
\epsilon_{ABCD} Y^A_l \Psi^{B\dagger}_l Y^C_l \Psi^{D\dagger}_l],
\end{align}
and the $V^{(1)}_{ferm}$ term become coincident in the scaling limit. Therefore, using the result in \cite{Honma:2008jd} that the ABJM fermionic
 potential scales as
\begin{align}
\Bar{\psi} X_0^I [X^J ,\Gamma_{IJ} \psi],
\end{align}
we can say that the scaling limit of the $V^{(1)}_{ferm}$ term is given by
\begin{align}
-\frac{\pi}{2k} \Bar{\psi}_{(m)} X_0^I [X^J ,\Gamma_{IJ} \psi]_{(-m)},
\end{align}
where $\psi^T_{(m)}=(\chi^T_{1(m)}, \cdots , \chi^T_{8(m)})$. This agrees with the second term of (\ref{ELfer}).

Therefore, we completely verify the emergence of the Extended L-BLG theory with two Lorentzian pairs from the scaling limit of the
 ${\cal{N}}=4$ quiver CS theory. This means that we obtain a concrete prescription for gaining D3-brane theory from the ABJM theory, because the
 Extended L-BLG theory with $d=1$ can be reduced to the D3-brane theory.
\section{Applications to the other quiver Chern-Simons theories}
Thus far, we have only discussed a particular ${\cal{N}}=4$ quiver CS theory (\ref{N4QCS}). However, by orbifolding the ABJM theory, we can
 obtain infinitely many quiver CS theories. Thus, here, we apply our scaling limit to various quiver CS theories.
\subsubsection*{(I) ${\mathbb{C}}^2 \times {\mathbb{C}}^2/{\mathbb{Z}}_{n}$}
\begin{figure}[h]
\centering
\includegraphics[width=14cm]{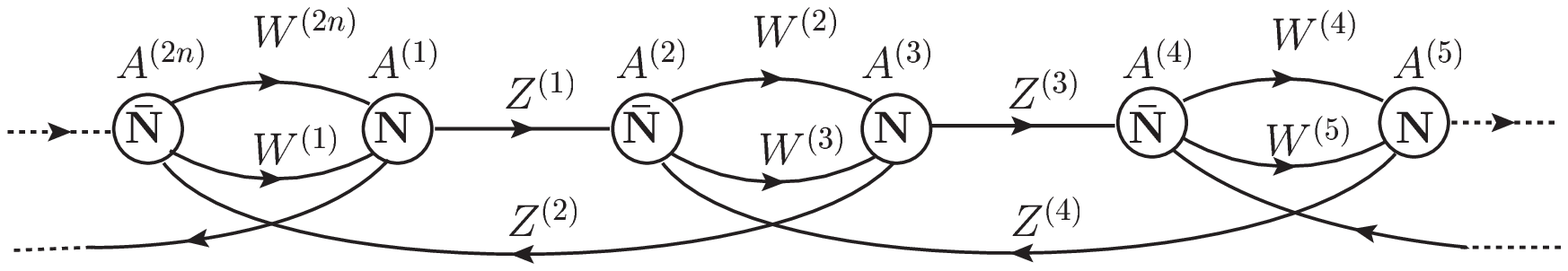}
\caption{Quiver diagram for case (I).}
\label{Fig2}
\end{figure}

The ${\mathbb{Z}}_n$ action (\ref{orb1}) was of the ${\mathbb{C}}^2 \times {\mathbb{C}}^2/{\mathbb{Z}}_{n}$ type. As another example of this
 type, let us consider the following ${\mathbb{Z}}_n$ orbifolding action\footnote{This is the ``chiral orbifold gauge theory" described in
 \cite{Benna:2008zy}.}:
\begin{align}
y^1 \rightarrow e^{2\pi i/n}y^1, \ \ y^2 \rightarrow e^{-2\pi i/n}y^2, \ \ y^3 \rightarrow y^3, \ \ y^4 \rightarrow y^4.
\end{align}
This preserves ${\cal{N}}=2$ supersymmetry and $SU(2)$ global symmetry.
 The covariant derivatives are 
\begin{align}
D_{\mu} Z^{(2l-1)} &=\partial_{\mu}Z^{(2l-1)}+iA^{(2l-1)}_{\mu}Z^{(2l-1)}-iZ^{(2l-1)}A^{(2l)}_{\mu}, \nonumber \\
D_{\mu} Z^{(2l)} &=\partial_{\mu}Z^{(2l)}+iA^{(2l+1)}_{\mu}Z^{(2l)}-iZ^{(2l)}A^{(2l-2)}_{\mu}, \nonumber \\
D_{\mu} W^{(2l-1)} &=\partial_{\mu}W^{(2l-1)}+iA^{(2l-2)}_{\mu}W^{(2l-1)}-iW^{(2l-1)}A^{(2l-1)}_{\mu}, \nonumber \\
D_{\mu} W^{(2l)} &=\partial_{\mu}W^{(2l)}+iA^{(2l)}_{\mu}W^{(2l)}-iW^{(2l)}A^{(2l+1)}_{\mu},
\end{align}
where $l=1,\cdots,n$.
 The $Z^{(2l)} , W^{(2l-1)}$ parts are changed from the ${\cal{N}}=4$ case (\ref{coder1}). Figure \ref{Fig2} is the corresponding quiver diagram.

In this theory, the Chern-Simons term is unchanged from the ${\cal{N}}=4$ case. Thus, its scaling limit is completely the same as that of (\ref{CS}).
As for the kinetic term, the covariant derivatives are scaled as
\begin{align}
&D_{\mu}Z_{(2l)} \rightarrow \frac{q^{lm}}{\sqrt{2n}}\partial_{\mu}Y^{(2)}_{(m)}+i\frac{q^{lm}}{\sqrt{2n}}[A_{\mu (n)},Y^{(2)}_{(m-n)}]-
\frac{2\pi (s+2) m q^{lm}}{\sqrt{2n}}A_{\mu (m)}Y^{(2)}_0+i\frac{2\pi q^{lm}}{\sqrt{2n}}A'_{\mu (m)}Y^{(2)}_0, \nonumber \\
&D_{\mu}W_{(2l-1)} \rightarrow \frac{q^{-lm}}{\sqrt{2n}}\partial_{\mu}Y^{(3) \dagger}_{(m)}+i\frac{q^{lm}}{\sqrt{2n}}[A_{\mu (n)},
Y^{(3) \dagger}_{(n-m)}]+\frac{2\pi (s+1) m q^{lm}}{\sqrt{2n}}A_{\mu (m)}Y^{(3) \dagger}_0 \nonumber \\
& \ \ \ \ \ \ \ \ \ \ \ \ \ \ \ \ \ \ -i\frac{2\pi q^{lm}}{\sqrt{2n}}A'_{\mu (m)}Y^{(3) \dagger}_0.
\end{align}
Again, through the assignments
\begin{align}
&\lambda^{I0}=-2\pi (X^1_0,X^2_0,X^3_0,X^4_0,X^5_0,X^6_0,X^7_0,X^8_0), \nonumber \\
&\lambda^{I1}=-2\pi \Big( sX^1_0,(s+2)X^2_0 ,(s+1)X^3_0 , (s+1)X^4_0 ,sX^5_0 ,(s+2)X^6_0 ,(s+1)X^7_0 ,(s+1)X^8_0 \Big) ,
\end{align}
we see that the kinetic term completely agrees with (\ref{ELkine}). The constraint for the metric of two-torus is calculated as
\begin{align}
G^{11}=-s(s+1)G^{00}+(2s+1)G^{01}+8\pi^2 [(X^2_0)^2+(X^6_0)^2].
\label{VEV2}
\end{align}
The difference from the previous case is an appearance of a term $(X^2_0)^2+(X^6_0)^2$.
 This indicates that we can cover a larger parameter space of the coupling constant $\tau$ than the ${\cal{N}}=4$ quiver CS
 theories, as we will see in Section 4.

\subsubsection*{(II) ${\mathbb{C}} \times {\mathbb{C}}^3/{\mathbb{Z}}_{n}$}
\begin{figure}
\centering
\includegraphics[width=16cm]{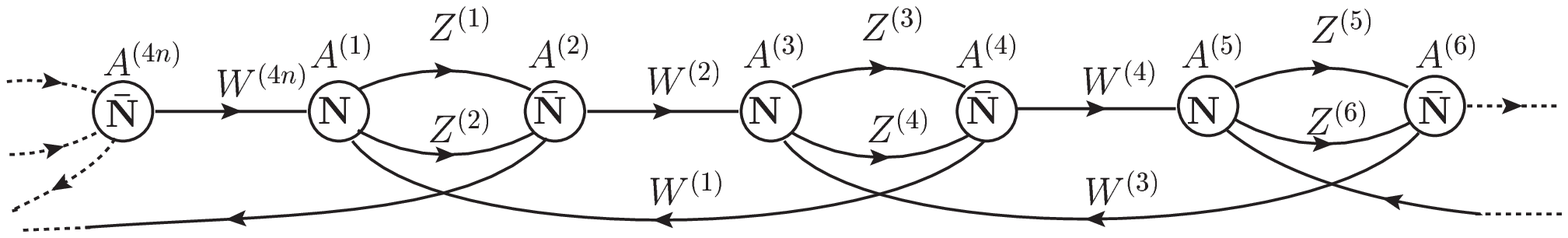}
\caption{Quiver diagram for case (II)-(i).}
\label{Fig3}
\end{figure}

\subparagraph{(i)}
Now, we consider the ${\mathbb{Z}}_{2n}$ action given by
\begin{align}
y^1 \rightarrow e^{2\pi i/2n}y^1, \ \ y^2 \rightarrow e^{2\pi i/2n}y^2, \ \ y^3 \rightarrow e^{2\pi i/n}y^3, \ \ y^4 \rightarrow y^4.
\end{align}
The quiver CS theory based on this orbifolding also has ${\cal{N}}=2$ SUSY and $SU(2)$ global symmetry.
 The quiver diagram of this theory is given in Figure \ref{Fig3}.
 The covariant derivatives are given by
\begin{align}
D_{\mu} Z^{(2l-1)} &=\partial_{\mu}Z^{(2l-1)}+iA^{(2l-1)}_{\mu}Z^{(2l-1)}-iZ^{(2l-1)}A^{(2l)}_{\mu}, \nonumber \\
D_{\mu} Z^{(2l)} &=\partial_{\mu}Z^{(2l)}+iA^{(2l-1)}_{\mu}Z^{(2l)}-iZ^{(2l)}A^{(2l)}_{\mu}, \nonumber \\
D_{\mu} W^{(2l-1)} &=\partial_{\mu}W^{(2l-1)}+iA^{(2l+2)}_{\mu}W^{(2l-1)}-iW^{(2l-1)}A^{(2l-1)}_{\mu}, \nonumber \\
D_{\mu} W^{(2l)} &=\partial_{\mu}W^{(2l)}+iA^{(2l)}_{\mu}W^{(2l)}-iW^{(2l)}A^{(2l+1)}_{\mu},
\end{align}
where $l=1,\cdots,2n$. The $Z^{(2l)} , W^{(2l-1)}$ parts are changed from (\ref{coder1}). The Chern-Simons term is unchanged from the one
 in (\ref{N4QCS}) except that $l$ runs $1$ to $2n$.

\begin{figure}
\centering
\includegraphics[width=12cm]{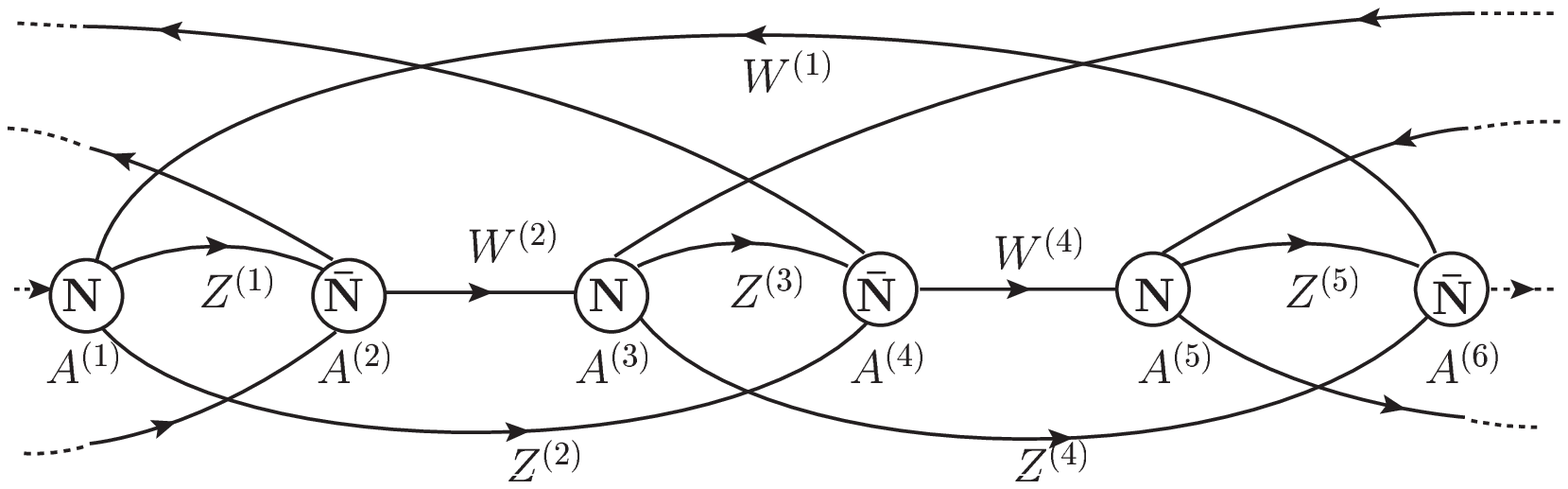}
\caption{Quiver diagram for case (II)-(ii).}
\label{Fig4}
\end{figure}

In this case, we have to change the scaling limit (\ref{limit}) slightly. Because we took a ${\mathbb{Z}}_{2n}$ orbifolding, we must change $n$ to
 $2n$ in (\ref{limit}) and redefine $q$ as $q \equiv e^{\frac{2\pi i}{2n}}$. Under this limit, the CS term of the Extended L-BLG theory is
 properly derived. The covariant derivatives are scaled as
\begin{align}
&D_{\mu}Z_{(2l)} \rightarrow \frac{q^{lm}}{\sqrt{4n}}\partial_{\mu}Y^{(2)}_{(m)}+i\frac{q^{lm}}{\sqrt{4n}}[A_{\mu (n)},Y^{(2)}_{(m-n)}]-
\frac{2\pi sm q^{lm}}{\sqrt{4n}}A_{\mu (m)}Y^{(2)}_0+i\frac{2\pi q^{lm}}{\sqrt{4n}}A'_{\mu (m)}Y^{(2)}_0, \nonumber \\
&D_{\mu}W_{(2l-1)} \rightarrow \frac{q^{-lm}}{\sqrt{4n}}\partial_{\mu}Y^{(3) \dagger}_{(m)}+i\frac{q^{lm}}{\sqrt{4n}}[A_{\mu (n)},
Y^{(3) \dagger}_{(n-m)}]+\frac{2\pi (s-1)m q^{lm}}{\sqrt{4n}}A_{\mu (m)}Y^{(3) \dagger}_0 \nonumber \\
& \ \ \ \ \ \ \ \ \ \ \ \ \ \ \ \ \ \ -i\frac{2\pi q^{lm}}{\sqrt{4n}}A'_{\mu (m)}Y^{(3) \dagger}_0.
\end{align}
Under the identifications
\begin{align}
&\lambda^{I0}=-2\pi (X^1_0,X^2_0,X^3_0,X^4_0,X^5_0,X^6_0,X^7_0,X^8_0), \nonumber \\
&\lambda^{I1}=-2\pi \Big( sX^1_0,sX^2_0 ,(s-1)X^3_0 , (s+1)X^4_0 ,sX^5_0 ,sX^6_0 ,(s-1)X^7_0 ,(s+1)X^8_0 \Big) ,
\end{align}
we can show the agreement of kinetic terms.
 The constraint to the $T^2$ metric is
\begin{align}
G^{11}=-s(s+1)G^{00}+(2s+1)G^{01}+8\pi^2 [(X^3_0)^2+(X^7_0)^2].
\label{VEV3}
\end{align}
Note that we have a degree of freedom that corresponds to tuning $[(X^3_0)^2+(X^7_0)^2]$ as with the case (I).
\subparagraph{(ii)}
Next, as another example of the ${\mathbb{C}} \times {\mathbb{C}}^3/{\mathbb{Z}}_{n}$ type, we consider the ${\mathbb{Z}}_{6n}$ action given
 by
\begin{align}
y^1 \rightarrow e^{2\pi i/6n}y^1, \ \ y^2 \rightarrow e^{2\pi i/3n}y^2, \ \ y^3 \rightarrow e^{2\pi i/2n}y^3, \ \ y^4 \rightarrow y^4.
\end{align}
This orbifold projection also preserves ${\cal{N}}=2$ SUSY, but the remaining global symmetry is less than before.
 The quiver CS theory obtained from this orbifold action has the following covariant derivatives,
\begin{align}
D_{\mu} Z^{(2l-1)} &=\partial_{\mu}Z^{(2l-1)}+iA^{(2l-1)}_{\mu}Z^{(2l-1)}-iZ^{(2l-1)}A^{(2l)}_{\mu}, \nonumber \\
D_{\mu} Z^{(2l)} &=\partial_{\mu}Z^{(2l)}+iA^{(2l-1)}_{\mu}Z^{(2l)}-iZ^{(2l)}A^{(2l+2)}_{\mu}, \nonumber \\
D_{\mu} W^{(2l-1)} &=\partial_{\mu}W^{(2l-1)}+iA^{(2l+4)}_{\mu}W^{(2l-1)}-iW^{(2l-1)}A^{(2l-1)}_{\mu}, \nonumber \\
D_{\mu} W^{(2l)} &=\partial_{\mu}W^{(2l)}+iA^{(2l)}_{\mu}W^{(2l)}-iW^{(2l)}A^{(2l+1)}_{\mu},
\end{align}
where $l=1,\cdots,6n$. Again, the $Z^{(2l)} , W^{(2l-1)}$ parts are changed from (\ref{coder1}). The corresponding quiver diagram is given in Figure \ref{Fig4}.

For the Chern-Simons term, under the scaling limit (\ref{limit}) with $n$ being replaced by $6n$, the agreement between both theories is easily
 shown as before. For the kinetic term, the covariant derivatives are scaled as
\begin{align}
&D_{\mu}Z_{(2l)} \rightarrow \frac{q^{lm}}{\sqrt{12n}}\partial_{\mu}Y^{(2)}_{(m)}+i\frac{q^{lm}}{\sqrt{12n}}[A_{\mu (n)},Y^{(2)}_{(m-n)}]-
\frac{2\pi (s-1)m q^{lm}}{\sqrt{12n}}A_{\mu (m)}Y^{(2)}_0+i\frac{2\pi q^{lm}}{\sqrt{12n}}A'_{\mu (m)}Y^{(2)}_0, \nonumber \\
&D_{\mu}W_{(2l-1)} \rightarrow \frac{q^{-lm}}{\sqrt{12n}}\partial_{\mu}Y^{(3) \dagger}_{(m)}+i\frac{q^{lm}}{\sqrt{12n}}[A_{\mu (n)},
Y^{(3) \dagger}_{(n-m)}]+\frac{2\pi (s-2)m q^{lm}}{\sqrt{12n}}A_{\mu (m)}Y^{(3) \dagger}_0 \nonumber \\
& \ \ \ \ \ \ \ \ \ \ \ \ \ \ \ \ \ \ -i\frac{2\pi q^{lm}}{\sqrt{12n}}A'_{\mu (m)}Y^{(3) \dagger}_0.
\end{align}
The agreement of kinetic terms is achieved using the assignment
\begin{align}
&\lambda^{I0}=-2\pi (X^1_0,X^2_0,X^3_0,X^4_0,X^5_0,X^6_0,X^7_0,X^8_0), \nonumber \\
&\lambda^{I1}=-2\pi \Big( sX^1_0,(s-1)X^2_0 ,(s-2)X^3_0 , (s+1)X^4_0 ,sX^5_0 ,(s-1)X^6_0 ,(s-2)X^7_0 ,(s+1)X^8_0 \Big).
\end{align}
In this case, the metric of $T^2$ is constrained to satisfy
\begin{align}
G^{11}=-s(s+1)G^{00}+(2s+1)G^{01}+8\pi^2 \{ (X^2_0)^2+(X^6_0)^2 \} +24\pi^2 \{ (X^3_0)^2+(X^7_0)^2 \}.
\label{VEV4}
\end{align}
Once again, we have a degree of freedom that corresponds to the sum of VEV squared.
\subsubsection*{(III) ${\mathbb{C}}^4/{\mathbb{Z}}_{n}$}
\begin{figure}
\centering
\includegraphics[width=13cm]{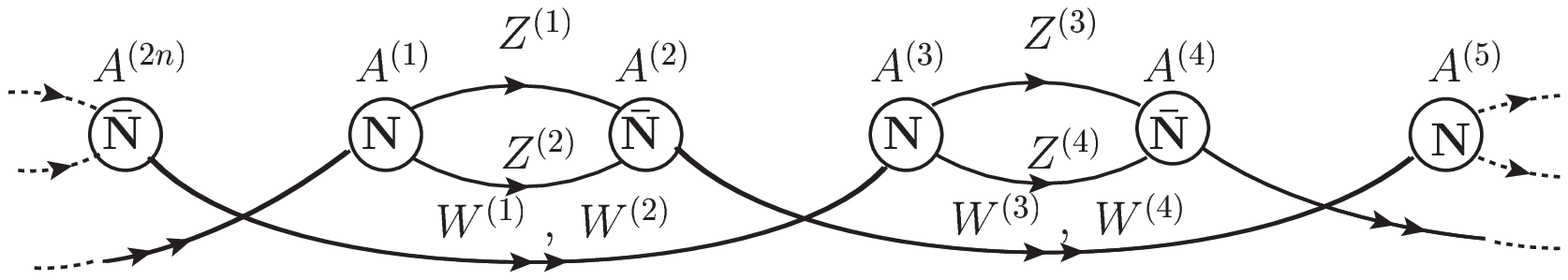}
\caption{Quiver diagram for case (III).}
\label{Fig5}
\end{figure}

Finally, we consider the ${\mathbb{C}}^4/{\mathbb{Z}}_{n}$ type.
When we consider the ${\mathbb{Z}}_{n}$ action given by
\begin{align}
y^1 \rightarrow e^{2\pi i/n}y^1, \ \ y^2 \rightarrow e^{2\pi i/n}y^2, \ \ y^3 \rightarrow e^{-2\pi i/n}y^3, \ \ y^4 \rightarrow e^{-2\pi
 i/n}y^4,
\end{align}
${\cal{N}}=4$ SUSY and $SU(2) \times SU(2)$ global symmetry are preserved. The covariant derivatives are given by
\begin{align}
D_{\mu} Z^{(2l-1)} &=\partial_{\mu}Z^{(2l-1)}+iA^{(2l-1)}_{\mu}Z^{(2l-1)}-iZ^{(2l-1)}A^{(2l)}_{\mu}, \nonumber \\
D_{\mu} Z^{(2l)} &=\partial_{\mu}Z^{(2l)}+iA^{(2l-1)}_{\mu}Z^{(2l)}-iZ^{(2l)}A^{(2l)}_{\mu}, \nonumber \\
D_{\mu} W^{(2l-1)} &=\partial_{\mu}W^{(2l-1)}+iA^{(2l-2)}_{\mu}W^{(2l-1)}-iW^{(2l-1)}A^{(2l+1)}_{\mu}, \nonumber \\
D_{\mu} W^{(2l)} &=\partial_{\mu}W^{(2l)}+iA^{(2l-2)}_{\mu}W^{(2l)}-iW^{(2l)}A^{(2l+1)}_{\mu},
\end{align}
where $l=1,\cdots,n$. In this case, only the $Z^{(2l-1)}$ part is unchanged from (\ref{coder1}).
 The quiver diagram of this theory is given in Figure \ref{Fig5}.

The CS term and its scaling behaviour are exactly the same as (\ref{N4QCS}) and (\ref{CS}), respectively. On the other hand, the covariant derivatives are scaled as
\begin{align}
&D_{\mu}Z_{(2l)} \rightarrow \frac{q^{lm}}{\sqrt{n}}\partial_{\mu}Y^{(2)}_{(m)}+i\frac{q^{lm}}{\sqrt{n}}[A_{\mu (n)},Y^{(2)}_{(m-n)}]-\frac{2\pi
 sm q^{lm}}{\sqrt{n}}A_{\mu (m)}Y^{(2)}_0+i\frac{2\pi q^{lm}}{\sqrt{n}}A'_{\mu (m)}Y^{(2)}_0, \nonumber \\
&D_{\mu}W_{(2l-1)} \rightarrow \frac{q^{-lm}}{\sqrt{n}}\partial_{\mu}Y^{(3) \dagger}_{(m)}+i\frac{q^{lm}}{\sqrt{n}}[A_{\mu (n)},Y^{(3) \dagger
}_{(n-m)}]+\frac{2\pi (s+2) m}{\sqrt{n}}q^{lm}A_{\mu (m)}Y^{(3) \dagger}_0-i\frac{2\pi q^{lm}}{\sqrt{n}}A'_{\mu (m)}Y^{(3) \dagger}_0, \nonumber \\
&D_{\mu}W_{(2l)} \rightarrow \frac{q^{-lm}}{\sqrt{n}}\partial_{\mu}Y^{(4) \dagger}_{(m)}+i\frac{q^{lm}}{\sqrt{n}}[A_{\mu (n)},Y^{(4) \dagger
}_{(n-m)}]+\frac{2\pi (s+2)m q^{lm}}{\sqrt{n}}A_{\mu (m)}Y^{(4) \dagger}_0-i\frac{2\pi q^{lm}}{\sqrt{n}}A'_{\mu (m)}Y^{(4) \dagger}_0.
\end{align}
Using the identifications
\begin{align}
&\lambda^{I0}=-2\pi (X^1_0,X^2_0,X^3_0,X^4_0,X^5_0,X^6_0,X^7_0,X^8_0), \nonumber \\
&\lambda^{I1}=-2\pi \Big( sX^1_0,sX^2_0 ,(s+2)X^3_0 , (s+2)X^4_0 ,sX^5_0 ,sX^6_0 ,(s+2)X^7_0 ,(s+2)X^8_0 \Big),
\end{align}
we can show that the kinetic term of the Extended L-BLG theory emerges precisely.
Therefore, the $T^2$ metric is limited to satisfy
\begin{align}
G^{11}=-s(s+2)G^{00}+(2s+2)G^{01}.
\label{VEV5}
\end{align}

In this section, we checked the emergence of the Extended L-BLG theory from the various quiver CS theories for the kinetic and
 CS terms. Naively, whenever an additional circle exists, independently of how to realize it, the Extended L-BLG theory and D3-brane theory
 are expected to emerge. Therefore, it is just conceivable that independently of how the further ${\mathbb{Z}}_n$ orbifolding acts on 
${\mathbb{C}}^4/{\mathbb{Z}}_k$, namely, regardless of the remaining SUSY and global symmetry, the orbifolded ABJM theories lead us to the
 Extended L-BLG theory from our scaling procedure. All the examples we have studied display positive signs for this expectation.
 Further research in this direction may be interesting.
\section{$T^2$ compactification and $SL(2,Z)$ transformations}
\setcounter{equation}{0}
We have seen the emergence of the Extended Lorentzian BLG theory from the scaling limit of quiver Chern-Simons theories.
 Our procedure realizes ordinary $T^2$ compactification.
  However, starting from the orbifolded ABJM theory, the resultant metric of two-torus $G^{AB} \ (A,B=0,1)$ is constrained. This means that after
 the reduction to  the D3-brane theory, the realizable parameter region of the complexified coupling constant $\tau$ is also limited. 
 In this section, we focus on this constraint and a realization of $SL(2,Z)$ transformations.

In section 2, we have seen that the Extended L-BLG theory with $d=1$ is reduced to the D3-brane worldvolume theory through the Higgs
 mechanism. The gauge sector of the resultant D3-brane action is given by
\begin{align}
L_A+L_{F\tilde{F}}&=-\frac{1}{4G^{00}}\int \frac{dy}{2\pi}\sqrt{g^{11}}F^2+\frac{G^{01}}{8G^{00}}\int \frac{dy}{2\pi}F\tilde{F} \nonumber \\
&\equiv -\frac{1}{8\pi}\int dy \left[ \textrm{Im}(\tau) F^2+\frac{1}{2}\textrm{Re}(\tau)F\tilde{F} \right],
\end{align}
where
\begin{align}
F^2&=\tilde{F}^2_{\mu\nu}+2g^{11}\tilde{F}_{\mu 1}\tilde{F}_{\mu 1}, \nonumber \\
F\tilde{F}&=(4\sqrt{g^{11}} \epsilon^{\mu\nu\lambda})\tilde{F}_{\mu 1}\tilde{F}_{\nu\lambda}.
\end{align}
Thus, the complexified coupling constant $\tau$ is represented as
\begin{align}
\tau=-\frac{G^{01}}{G^{00}}+i\sqrt{\frac{G^{11}}{G^{00}}-\left( \frac{G^{01}}{G^{00}} \right)^2}.
\label{tau}
\end{align}
Note that we have chosen $k=1$. 

In the previous section, we have seen that the $T^2$ metric $G^{AB}$ is constrained to satisfy a certain relation.
 Now, we substitute these constraints into (\ref{tau}) and investigate the parameter space of $\tau$ and the $SL(2,Z)$ transformations.
\subsubsection*{(I) ${\cal{N}}=4$}
First, we consider the ${\cal{N}}=4$ case.
 Substituting (\ref{VEV1}) into (\ref{tau}), we obtain
\begin{align}
\tau =-\frac{G^{01}}{G^{00}}+i\sqrt{-\left( \frac{G^{01}}{G^{00}}-s \right) \left[ \frac{G^{01}}{G^{00}}-(s+1) \right] },
\end{align}
where 
\begin{align}
\frac{G^{01}}{G^{00}}=s+\frac{(X^2_0)^2+(X^4_0)^2+(X^6_0)^2+(X^8_0)^2}{(X^I_0)^2}.
\end{align}
This denotes that in a fixed $s$, namely, in certain linear combinations of the gauge fields (\ref{gaugeLC}), the realizable parameter space of $\tau$ is
 limited to the one that depends on only one real parameter, the ratio of the VEVs $G^{01}/G^{00}$.
 Remarkably, $s$ appears in $\tau$ only through the real part.
 When we shift $s$ as $s \rightarrow s+a \ (a \in {\mathbb{Z}})$, $\tau$ changes as $\tau \rightarrow \tau+a$.
 Therefore, the linear combinations of the gauge fields and the T-transformations have one-to-one correspondence. This is an extension of the work
 in \cite{Hashimoto:2008ij}. This correspondence also works in all the other examples (I), (II), (III) in Section 4.

If we define $\tau \equiv x+iy$, the realizable region of the coupling $\tau$ is represented as
\begin{align}
\left( x+\frac{2s+1}{2} \right)^2+y^2=\frac{1}{4}.
\end{align}
This is an upper part of a circle of radius $1/2$ whose center depends on the combinations of gauge fields.
 
Similarly, if we consider the constraint (\ref{VEV5}), the realizable parameter space of $\tau$ is represented as
\begin{align}
\left( x+s+1 \right)^2+y^2=1.
\end{align}
Again, $\tau$ becomes a one parameter curve.

 In both cases, even if we move all the values of VEVs $X^I_0$ and indices $s \ (s \in {\mathbb{Z}})$, we cannot cover the full parameter space of
 the complex structure moduli $\tau$.
\subsubsection*{(II) ${\cal{N}}=2$}
In the ${\cal{N}}=2$ case, the situation slightly changes. Now, $\tau$ is represented as
\begin{align}
\tau =-\frac{G^{01}}{G^{00}}+i\sqrt{-\left( \frac{G^{01}}{G^{00}}-s \right) \left[ \frac{G^{01}}{G^{00}}-(s+1) \right] +A} \ ,
\end{align}
where
\begin{align}
A \equiv \begin{cases}
              8\pi^2[(X^2_0)^2+(X^6_0)^2]/G^{00} & \text{for (\ref{VEV2})}, \\
              8\pi^2[(X^3_0)^2+(X^7_0)^2]/G^{00} & \text{for (\ref{VEV3})}, \\
              [8\pi^2\{ (X^2_0)^2+(X^6_0)^2\} +24\pi^2\{ (X^3_0)^2+(X^7_0)^2\} ]/G^{00} & \text{for (\ref{VEV4})}.
             \end{cases}
\end{align}
Now, owing to the existence of the term $A$, we can move a larger region of the complex structure $\tau$ than in the ${\cal{N}}=4$ case.
 The realizable region of $\tau$ is represented as
\begin{align}
\left( x+\frac{2s+1}{2} \right)^2+y^2=\frac{1}{4}+A.
\end{align}
Compared with case (I), we can change a radius of a circle by tuning $A$.
 Therefore, moving all the values of allowed $x \ (=-G^{01}/G^{00})$, $s \ (s \in {\mathbb{Z}})$, and $A$, we can realize the parameter space of
 $\tau$ more widely. Hence, it seems that the one parameter dependence of $\tau$ in the previous case is the reflection of the fact that
 3d ${\cal{N}}=4$ SUSY is very restricted.

 Finally, we comment on the $A$ term. Because $A$ is bounded above, again the whole region of the complex structure moduli cannot be
 reproduced. Naively, even if we consider the ${\mathbb{Z}}_n$ action that preserves no supersymmetry, the situation seems to be unchanged.
 This is slightly mysterious and more work is required.
\section{Conclusion and discussion}
\setcounter{equation}{0}
In this paper, we have explicitly shown that the BLG theory with two Lorentzian pairs is derived by taking a scaling limit of an ${\cal{N}}=4$ quiver
 Chern-Simons theory, which is obtained by orbifolding the ABJM action.
 In this scaling limit, the VEVs are taken to be large compared with the fluctuating traceless components.
 Therefore, M2-branes are located far from the origin of the orbifold ${\mathbb{C}}^4/({\mathbb{Z}}_{kn} \times {\mathbb{Z}}_n)$. Then,
 taking $n \rightarrow \infty$ simultaneously, we effectively realize a standard $T^2$ compactification. This is why the Extended L-BLG theory
 emerges.

 Since the Extended L-BLG theory can be reduced to the Dp-brane worldvolume theory
 via the Higgs mechanism, our scaling procedure has useful applications for deriving Dp-branes from the ABJM theory.
 In this paper, we consider only the D3-brane case.
 We also investigate the scaling limit of various quiver CS theories and confirm that the kinetic and CS terms of the Extended L-BLG
 theory correctly emerge. Remarkably, it is found that the resulting D3-brane theory covers a larger region in the parameter space of the coupling
 constant $\tau$ than in the ${\cal{N}}=4$ case. In both cases, however, we cannot realize an entire region of the complex structure moduli.
 Naively, this situation seems to be unchanged even if we consider the non-SUSY case. This is slightly mysterious and more work is required.

There are some directions for further generalizations of this work. One direction is to understand the $d \geq 2$ case. Although we consider
 only the $d=1$ case in this paper, it seems that the more we orbifold the ABJM theory, the higher dimensional D-brane theory can be obtained.
 Moreover, it is just conceivable that independently of how ${\mathbb{Z}}_n$ orbifolding acts on ${\mathbb{C}}^4/{\mathbb{Z}}_k$, the
 orbifolded ABJM theory might lead to the Extended L-BLG theory (and Dp-brane theory via the Higgs mechanism) through our scaling procedure.
 Because the Extended L-BLG theory does not succeed in explaining several background fields in the $d \geq 2$ case, the understanding from the
 ABJM side may shed light on this problem. The generalization to M2-branes on general background is also interesting.
\section*{Acknowledgements}
We would like to thank S. Iso and Y. Orikasa for useful conversations. The work of S.Z. is supported in part by the JSPS Research Fellowship for
Young Scientists.
\appendix
\section{Gamma Matrices}
The explicit forms of the antisymmetrized $\Gamma$ matrices that we used in Section 3 are
\begin{align}
 &\Gamma_{12} = \left(
 \begin{array}{@{\,}cccc@{\,}}
  -i\sigma^2 &&&\\&i\sigma^2&&\\
  &&-i\sigma^2&\\&&&-i\sigma^2
 \end{array}
 \right), \quad
 &\Gamma_{13} &= \left(
 \begin{array}{@{\,}cccc@{\,}}
  &-\mathbb{I}&&\\ \mathbb{I}&&&\\
  &&&-\sigma^3\\&&\sigma^3&
 \end{array}\right),\notag\\ 
 &\Gamma_{14} = \left(
 \begin{array}{@{\,}cccc@{\,}}
  &-i\sigma^2&&\\-i\sigma^2&&&\\
  &&&-\sigma^1\\&&\sigma^1&
 \end{array}
 \right),\quad
 &\Gamma_{15} &= \left(
 \begin{array}{@{\,}cccc@{\,}}
  &&\sigma^3&\\&&&-\mathbb{I}\\
  -\sigma^3&&&\\&\mathbb{I}&&
 \end{array}\right),\notag\\ 
 &\Gamma_{16}=\left(
 \begin{array}{@{\,}cccc@{\,}}
  &&\sigma^1&\\&&&i\sigma^2\\
  -\sigma^1&&&\\&i\sigma^2
 \end{array}
 \right),\quad
 &\Gamma_{17} &= \left(
 \begin{array}{@{\,}cccc@{\,}}
  &&&\sigma^3\\&&\mathbb{I}&\\
  &-\mathbb{I}&&\\-\sigma^3&&&
 \end{array}\right),\notag\\ 
 &\Gamma_{18} = \left(
 \begin{array}{@{\,}cccc@{\,}}
  &&&\sigma^1\\&&-i\sigma^2\\
  &-i\sigma^2&&\\-\sigma^1&&&
 \end{array}\right)\quad,
 &\Gamma_{52} &= \left(
 \begin{array}{@{\,}cccc@{\,}}
  &&-\sigma^1&\\&&&i\sigma^2\\
  \sigma^1&&&\\&i\sigma^2&&
 \end{array}\right),\notag\\ 
 &\Gamma_{53}=\left(
 \begin{array}{@{\,}cccc@{\,}}
  &&&-\mathbb{I}\\&&-\sigma^3\\
  &\sigma^3&&\\ \mathbb{I}&&&
 \end{array}\right),\quad
 &\Gamma_{54}&=\left(
 \begin{array}{@{\,}cccc@{\,}}
  &&&-i\sigma^2\\&&-\sigma^1&\\
  &\sigma^1&&\\-i\sigma^2&&&
 \end{array}\right),\notag\\ 
 &\Gamma_{56}=\left(
 \begin{array}{@{\,}cccc@{\,}}
  -i\sigma^2&&&\\&-i\sigma^2&&\\
  &&-i\sigma^2&\\&&&i\sigma^2\\
 \end{array}\right),\quad
 &\Gamma_{57}&=\left(
 \begin{array}{@{\,}cccc@{\,}}
  &-\sigma^3&&\\\sigma^3&&&\\
  &&&-\mathbb{I}\\&&\mathbb{I}&
 \end{array}\right),\notag\\ 
 &\Gamma_{58}=\left(
 \begin{array}{@{\,}cccc@{\,}}
  &-\sigma^1&&\\\sigma^1&&&\\
  &&&-i\sigma^2\\&&-i\sigma^2&
 \end{array}\right) .
 \label{gamma matrices}
\end{align}
They indeed satisfy the consistency conditions as
$\Gamma_{12} \Gamma_{13}+\Gamma_{13} \Gamma_{12} = -(\Gamma_2\Gamma_3 +
\Gamma_3\Gamma_2) = 0$.

\end{document}